\documentclass[single,oneside]{araa_jsb}
\usepackage[]{natbib}

\usepackage{psfig,epsfig}	

\def\gtaprx {\lower .1ex\hbox{\rlap{\raise .6ex\hbox{\hskip .3ex
	{\ifmmode{\scriptscriptstyle >}\else
		{$\scriptscriptstyle >$}\fi}}}
	\kern -.4ex{\ifmmode{\scriptscriptstyle \sim}\else
		{$\scriptscriptstyle\sim$}\fi}}}
\def\ltaprx {\lower .1ex\hbox{\rlap{\raise .6ex\hbox{\hskip .3ex
	{\ifmmode{\scriptscriptstyle <}\else
		{$\scriptscriptstyle <$}\fi}}}
	\kern -.4ex{\ifmmode{\scriptscriptstyle \sim}\else
		{$\scriptscriptstyle\sim$}\fi}}}
\newcommand{\Msun}{M$_{\odot}$}

\newcommand{\citepeg}[1]{\citep[{e.g.,}][]{#1}}

\def\lsim{\hbox{ \rlap{\raise 0.425ex\hbox{$<$}}\lower 0.65ex\hbox{$\sim$}}}
\def\gsim{\hbox{ \rlap{\raise 0.425ex\hbox{$>$}}\lower 0.65ex\hbox{$\sim$}}}
\def\lt{\hbox{ \rlap{\raise 0.425ex\hbox{$<$}}\lower 0.65ex\hbox{$\sim$}}}
\def\gt{\hbox{ \rlap{\raise 0.425ex\hbox{$>$}}\lower 0.65ex\hbox{$\sim$}}}

\def\ale{\mathrel{\hbox{\rlap{\hbox{\lower4pt\hbox{$\sim$}}}\hbox{$<$}}}}
\def\age{\mathrel{\hbox{\rlap{\hbox{\lower4pt\hbox{$\sim$}}}\hbox{$>$}}}}

\newcommand{\lSect}[1]{{\label{sec:#1}}}

\newcommand{\Sectff}[1]{{\ref{sec:#1}}}
\newcommand{\Sect}[1]{{\S~\Sectff{#1}}}

\newcommand{\iaucirc}{{\sl IAU Circ}}

\newcommand{\aj}{{\sl Astronomical Journal}}

\newcommand{\apjl}{{\sl Ap. J. Lettr.}}
\newcommand{\aap}{{\sl Astron. and Ap.}}
\newcommand{\apj}{{\sl Ap. J.}}
\newcommand{\apjs}{{\sl Ap. J. Suppl.}}
\newcommand{\mnras}{{\sl MNRAS}}
\newcommand{\prd}{{\sl Phys. Rev. D}}
\newcommand{\apss}{{\sl Ap. and Spac. Sci.}}
\newcommand{\nat}{{\sl Nature}}
\newcommand{\physrep}{{\sl Physics Reports}}
\newcommand{\araa}{{\sl Ann. Rev. Astron. Ap.}}
\newcommand{\pasp}{{\sl Proc. Ast. Society Pac.}}

\newcommand\ion[2]{#1$\;${\small #2}\relax}%

\begin{document}

\jname{Annual Rev. Astron. Astrophys.}
\jyear{2006}
\jvol{44}
\ARinfo{10.1146/annurev.astro.43.072103.150558}

\title{The Supernova Gamma-Ray Burst Connection}

\markboth{Woosley \& Bloom}{The Supernova Gamma-Ray Burst Connection}

\author{S. E. Woosley
\affiliation{Department of Astronomy and Astrophysics, UCSC, Santa Cruz, CA 95064\\
woosley@ucolick.org}
J. S. Bloom
\affiliation{Department of Astronomy, University of California, Berkeley, CA 94720\\
jbloom@astro.berkeley.edu}}


\begin{abstract}
Observations show that at least some gamma-ray bursts (GRBs) happen
simultaneously with core-collapse supernovae (SNe), thus linking by a
common thread nature's two grandest explosions. We review here the
growing evidence for and theoretical implications of this association,
and conclude that most long-duration soft-spectrum GRBs are
accompanied by massive stellar explosions (GRB-SNe). The kinetic
energy and luminosity of well-studied GRB-SNe appear to be
greater than those of ordinary SNe, but evidence exists, even in a
limited sample, for considerable diversity. The existing sample also
suggests that most of the energy in the explosion is contained in
nonrelativistic ejecta (producing the supernova) rather than in the
relativistic jets responsible for making the burst and its
afterglow. Neither all SNe, nor even all SNe of Type Ibc produce GRBs.
The degree of differential rotation in the collapsing iron core of
massive stars when they die may be what makes the difference.
\end{abstract}

\maketitle

\section{INTRODUCTION}
\label{intro}

Gamma-ray bursts (GRBs), discovered by \citet{kso73}, are brief
($\sim$seconds), intense flashes of electromagnetic radiation with
typical photon energies $\sim$100 keV that arrive at Earth from
unpredictable locations several times daily \citep[e.g.,][]{Fis95}.
They are isotropically distributed on the sky and, so far as we know,
none has ever repeated\footnote{Not included in our review here are
the ``soft-gamma repeaters''
\citep{wt+05}, related phenomenologically to classic GRBs but
believed to be associated with highly magnetic neutron stars in the
local group of galaxies
\citep[though see][]{Tan05}. There is no known direct temporal
connection of SGRs to supernovae.}. The production of GRBs is believed
to require some small amount of matter accelerated to
ultra-relativistic speeds \citep[e.g.,][]{Mes02} and beamed to a small
fraction of the sky \citep{fks+01}. In many of the longer lasting
events the total energy in gamma-rays, corrected for this beaming, is
$\sim$10$^{51}$ erg.

Core-collapse supernovae (SNe), on the other hand, are the explosive
deaths of massive stars that occur when their iron cores collapse to
neutron stars or black holes \citep[e.g.,][]{Woo05b}.  They are, in
general, not accompanied by highly relativistic mass ejection, but are
visible from all angles and last from weeks to months.  They may be
either of Type Ib or Ic, if their hydrogen envelopes are lost, or Type
II if they are not \citep{fil97}. The total {\sl kinetic} energy here
is also $\sim$10$^{51}$ erg, roughly the same as the energy of the jet
that makes a GRB.

SNe are the most {\sl powerful} explosions in the modern universe,
rivaling in pure wattage the rest of the observable universe combined,
but most of the emission, $\sim$10$^{53}$ erg s$^{-1}$, is in
neutrinos which are, unless one happens to be very nearby,
unobservable. GRBs are the {\sl brightest} explosions in the universe,
in terms of electromagnetic radiation per unit solid angle, sometimes
as bright as if the rest mass of the sun, $2 \times 10^{54}$ erg, had
been turned to $\gamma$-rays in only ten seconds. The light per solid
angle from SNe is about ten orders of magnitude fainter.

Despite the similarity in kinetic energy scale, it was thought for
decades that these two phenomena had no relation to one another
\citep[though see][]{col68,pac86}. This was largely because, until the
late 1990's, no one knew just how far away --- and hence how luminous
--- the GRBs really were. Indeed, the consensus in the mid 1990's was
that even if GRBs were ever found to be of cosmological origin, they
were likely the result of a merger in a degenerate binary system, such
as double neutron stars \citep{Bli84,pac86,Eic89,npp92} or a neutron
star and black hole \citep{nps91,Pac91} and would not be found in
conjunction with supernovae.

Beginning in 1997 with the localization of long-wavelength
counterparts \citep{cfh+97,vgg+97} and the confirmation of the
cosmological distance scale \citep{mdk+97}\footnote{By 1997, most of
the community had already come to accept a cosmological distance scale
for GRBs because of the isotropic distribution of locations found by
the Burst and Transient Source Experiment \citep{Fis95,pac95}.}, it
became increasingly clear that those well-studied GRBs were associated
with young stars in distant actively star-forming galaxies, and not with
old stars in mature galaxies, as expected of the merger
hypothesis. Within a few years, evidence mounted against the merger
hypothesis and implicated GRBs, to the surprise of many, as due to the
death of massive stars (\Sect{hosts}).

This association with star formation did not require a causal
connection though. Perhaps the young stars resulted in SNe that
produced neutron stars or black holes which in turn made the bursts
some time later. Nor was it clear that any supernova accompanying a
GRB would be especially bright\footnote{Here we distinguish
``supernova,'' the sub-relativistic explosion of a stellar mass
object, from the ``optical afterglow'' produced when the relativistic
material responsible for a GRB impacts the surrounding medium
\citep{mr97a,Pir05}.}.  The watershed event that brought 
the SN--GRB connection to the forefront was the discovery of a GRB on
April 25, 1998 (GRB\,980425) in conjunction with one of the most
unusual SNe ever seen, SN\, 1998bw \citep{gvv+98}. The SN and the GRB
were coincident both in time and place. Theory had anticipated that
event \citep{Woo93}, but not its brilliance. Though the physical
connection was initially doubted by some, and GRB\, 980425 was very
sub-energetic compared with most GRBs, the in-depth study of the
``SN-GRB connection'' began in earnest on that day (\Sect{sn1998bw}).

Aspects of the SN-GRB connection have been reviewed previously
\citep{ww98,vkw00,vp01,wmh01,whe01,Wei02,Mes02,blo05,Pir05,math05,dv05,hbk+05}, 
but usually as cursory overviews of the connection, or parts of larger
reviews of GRBs in general. Here we review specifically the observations,
history, and theory relating to the SN-GRB connection.  Since, until
recently, the only accurately known counterparts were for GRBs of the
so called long-soft variety\footnote{Though the light curves and
spectra of cosmic GRBs are very diverse, they can be broadly
categorized on the basis of duration and power spectrum into two
groups
\citep{Kou93,Fis95} --- ``short-hard'' bursts with a median duration
of 0.3 s, and ``long-soft'' bursts with a median duration of 20 s. The
average peak energy of the shorter class is about 50\% greater than
the long class - 360 keV vs 220 keV \citep{Hak00}.}, our discussion
centers on these events (though see \Sect{shorthard} and \Sect{xrf}).

What we know now is that at least one other supernova besides SN
1998bw, namely SN 2003dh (\S \ref{sec:spect}), has happened nearly
simultaneously with a GRB. This time the GRB (030329) was of a more
normal energy. A compelling spectroscopic case can also be made for
supernova associations with GRB 031203 (SN 2003lw) and possibly GRB
021211 (SN 2002lt). There have also been ``bumps'' in the optical
afterglows of many GRBs (\Sect{bumps}) consistent in color, timing and
brightness with what is expected from Type I supernovae of luminosity
comparable to SN 1998bw. Indeed, given the difficulties in making the
key observations, the data and models we shall review
(\Sect{Observations}, \Sect{characteristics}, and \Sect{allsn}) are
consistent with, though not conclusive proof of, the hypothesis that
{\sl all} long-soft GRBs are accompanied by SNe of Type Ic
\footnote{A Type Ic supernova has no hydrogen in
its spectrum and also lacks strong lines of \ion{Si}{II} or
\ion{He}{I} that would make it Type Ib or Ia. See \citet{fil97} for
review.}. Still, there is evidence for considerable diversity in the
brightness, rise times, and evolution of these events. The
well-studied SNe that accompany GRBs (GRB-SNe) also show evidence for
{\sl broad lines}, indicative of high velocity ejecta. This leads us
to suggest a sub-classification of Type Ic SNe, called ``Type Ic-BL''
(whether they are associated with GRBs or not). Type Ic-BL is a purely
observational designation, and makes no reference to a specific
progenitor model (e.g., ``collapsars'') nor to the model-specific
energetics or brightness of the explosions. In the latter respect, the
GRB modeling community has come to view the label of ``hypernova'' for
typing GRB-SNe as somewhat narrow and subjective.

While all GRBs of the long-soft variety may be accompanied by SNe, not
all SNe, or even all SNe of Type Ic-BL, make GRBs. But why
should some stars follow one path to death and others, another?  As
described in
\Sect{progenitor}, {\sl rotation} is emerging as the
distinguishing ingredient. Though it is a conjecture still to be
proven, GRBs may only come from the most rapidly rotating and most
massive stars, possibly favored in regions of low
metallicity. Ordinary SNe, on the other hand, which comprise about
99\% of massive star deaths, may come from stars where rotation plays
a smaller role or no role at all. Indeed, the SN-GRB connection is
forcing a re-evaluation of the role of rotation in the deaths of all
sorts of massive stars.

The continued observation of both GRBs and the SNe that accompany them
should yield additional diagnostics that will help the community gain deeper
insight in to both phenomena. Some of these diagnostics, especially
those that might shed light on the prime mover, or ``central engine''
that drives all these explosions are discussed in \Sect{diagnostics},
and we end with a discussion of future directions in the field in
\Sect{future}.

\section{OBSERVATIONS}

\subsection{OBSERVATIONAL EVIDENCE FOR A SN-GRB ASSOCIATION}
\label{sec:Observations}

\subsubsection{Early Indications}

\citet{col68}, in the only paper to {\sl predict} the existence
of GRBs before their discovery, associated them with the breakout of
relativistic shocks from the surfaces of SNe. This motivated the
discoverers \citep{kso73} to search for SN-GRB coincidences, but none
was found. We know now that the transients from breakout itself are
too faint to be GRBs at cosmological distances\footnote{Aspects of the
Colgate model remain viable.  A relativistic blast wave interacting
with a circumstellar medium might be a reasonable model for some
events \citep{Woo99,Mat99,Tan01}, especially if the ejecta are
beamed.}. Even with thousands more GRBs and hundreds of SNe localized
by 1997, no clear observational connection could be established
\citepeg{hw88}.

Bohdan Paczy\'nski, for years (e.g., \citealt{pac86}; though see also
\citealt{uc75}), suggested a cosmological origin for GRBs, and pointed out
that the requisite energy (in $\gamma$-rays) would be comparable to
the (kinetic) energy of a supernova. When the first redshifts of GRBs
were determined \citep{mdk+97,kdr+98}, the implied energies were up to
$\sim$10$^3$ times larger than $10^{51}$ erg. In fact, the largest
inferred $\gamma$-ray energy exceeded the rest mass energy of a
neutron star \citep{kdo+99}. Later, however, with geometric
corrections for beaming \citep{rho97b,hum+00}, the total energy
release in $\gamma$-rays came down to around 10$^{51}$ erg
\citep{kp00a,fks+01,fw01,bfk03,fb05}, with a small, but significant
number of bursts at lower energies \citepeg{skb+04}. That the actual
energy release in long-duration GRBs and SNe is comparable is
consistent with an association, but does not require a common origin.
Similar energetics are expected from a variety of viable cosmological
progenitors. Yet, even before direct confirmation, several independent
lines of evidence pointed tantalizingly to a direct SN-GRB connection. On
length scales spanning parsecs (``circumburst''), to galaxies, to cosmic
distances, GRBs revealed their origin.

\subsubsection{Location In and Around Distant Galaxies}
\lSect{hosts}

The various scenarios for making GRBs (\Sect{grbmodels}) have
implications for their observed locations. Since neutron stars experience
``kicks'' at birth, the long delay before coalescence would lead to
bursts farther from star-forming regions
\citep{pac98b,lsp+98,bsp99,fwh99,bbz00} than very massive stars. 
Therefore, subarcsecond localizations of afterglows with respect to
distant galaxies provided, early on, an indirect means for testing
hypotheses about the progenitors. GRB\, 970228, for example, was
localized on the outskirts of a faint galaxy, \citep{vgg+97}
essentially ruling out
\citep{slp+97} disruptive events around a central massive black hole
\citep{car92}. Unfortunately, the imaging capabilities of current
instruments cannot resolve the immediate environment ($\ale 100$ pc)
of GRBs that originate beyond $\sim$100 Mpc. Hence the location of
individual bursts (for instance, in an H II region) cannot be used as
a definitive test of their nature. However, statistical studies reveal
a strong correlation of the locations with the blue light of galaxies
\citep{bkd02,fru+05}. It seems that long-soft GRBs
happen preferentially in the regions where the most massive stars die.

No host galaxy stands out as exceptional, yet, in the aggregate, they
are faint and blue \citep{mm98,ldm+03}, systematically smaller,
dimmer, and more irregular than $M_*$ galaxies at comparable redshifts
\citep{mm98,hf99,dfk+03,cvf+05,wbp05}. Of the more than 60 GRB hosts known,
only one (GRB\,990705) appears to be associated with a normal spiral
\citep{mpp+00}, and that could be a coincidence (given the
large solid angle of a big face-on spiral, the GRB has a greater
probability of having occurred at higher redshift than its putative
host). No long-duration GRB has ever been definitively associated with
an early-type galaxy\footnotemark\footnotetext{One group's photometry
of the host of GRB\,970508 \citep{cba02} indicated a significant old
population, but this was not confirmed by other groups.  Furthermore,
morphological analysis showed that the galaxy surface brightness fit
better an exponential than an r$^{-1/4}$ profile \citep{fpg+00}.}.

There is convincing spectroscopic evidence that typical GRB host
galaxies are forming stars actively, perhaps at a higher rate per unit
mass than field galaxies \citep{dkb+01,chg04}. Sub-mm observations of
some GRB hosts once appeared to indicate the presence of large amounts
($\sim$hundreds $M_\odot$ yr$^{-1}$) of obscured star-formation
\citep{bck+03}, but recent mid-infrared observations of the same
galaxies suggest significantly smaller star-formation rates
\citep{fcf+06}. There are indications \citepeg{bdk01}, in particular
from the line ratios of [Ne III] to [OII], that the H II regions are
especially hot, indicating a propensity of GRB hosts to make more
massive stars. More recently, a growing body of absorption-line
spectroscopic evidence suggests that GRB hosts --- or more precisely,
the regions through which GRB afterglows are viewed --- are low in
metallicity \citep{vfk+01,sff03,pbc+04} though it remains to be seen
whether metallicities are significantly different from field galaxies
and damped Lyman $\alpha$ systems at comparable redshifts. As a class,
the hosts of long-duration GRBs thus appear to favor progenitors that
are closely connected with metal-poor massive stars (see
\Sect{metallicity}).


The locations of long-duration GRBs on the largest scale --- in
redshift space --- confirm the expectations from smaller spatial
scales. The GRB rate appears to track the global star-formation rate
\citep{lw98,blo03a,fag+04,ps04,nah+05,jlf+05}. This too
implicates a progenitor that makes a GRB without appreciable
($\age$Gyr) delay following a starburst
\citep{tot97,wbbn98}. While bursts from massive stars could, in
principle, be observable to redshifts of $\sim$30
\citep{mr03}, bursts from merger remnants with long delay 
times since starburst (such as, perhaps, short-hard GRBs) should not
be observed beyond $z \approx 6$: there would simply not be enough
time since the formation of the first stars.

\subsubsection{Absorption Spectroscopy}

Absorption line spectroscopy of GRB afterglows can also constrain the
environments around GRB progenitors \citep{pl98a}, on both galactic
and stellar scales.  The columns of neutral hydrogen and metals seen
in absorption in the highest redshift GRBs are generally significantly
larger than those seen through quasar sight lines
\citepeg{sff03}. This is not necessarily an effect local to the GRB,
but possibly a consequence of the location of GRBs in the inner
regions of their host.

Spectroscopy of the afterglow of GRB\,021004 revealed  significantly
blueshifted ($\age 100$ km s$^{-1}$) absorption features relative to the highest $z$ system \citep{cf02,mfh+02,vel+04}. While some have claimed that the high velocities could be due to radiative
acceleration of the circumburst material \citepeg{sgh+03}, a more natural explanation appears to be that the absorption is due to fast moving Wolf-Rayet winds from the progenitor. Indeed detailed
modeling of Wolf-Rayet winds and interactions with the interstellar
medium appear to accommodate significant column densities at a variety
of blueshifts extending out to $\sim$ 2000 km/s (van Marle, Langer, \&
Garcia Segura 2005; although see Mirabal {\sl et al.}
2002). High-resolution spectra of more recent bursts also appear to
show WR features (e.g., \citealt{pcb06}). The presence of fine
structure lines of, for example, Fe$^+$, indicates a warm, dense
medium that has only been observed in Galactic WR winds and never
been seen in any extragalactic sightlines. Interestingly, because of
the redshifting of UV lines that would be otherwise inaccessible to
groundbased spectroscopy, high resolution spectra of GRBs are now
offering unique detailed diagnostics of WR winds. Circumburst
diagnostics are discussed further in \Sect{spectrumag}.
\nocite{Lan05}\nocite{mhk+02}

\subsubsection{GRB\, 980425 and SN\, 1998bw}
\lSect{sn1998bw}

GRB\, 980425 triggered detectors on board both BeppoSAX and BATSE
\citep{kip98}. At high energies, it was seemingly unremarkable
\citep{kip98,gvv+98} with a typical soft spectrum ($E_{\rm peak}
\approx 150$ keV) and moderate duration ($\Delta T \approx 23$ sec).
Within the initial 8 arcmin radius error circle was an underluminous
(0.02 $L_*$) late-type galaxy (ESO184$-$G82; $z = 0.0085$, 
\citealt{tsc+98}), found 2.5 days after the GRB to host a young
supernova, designated as SN 1998bw
\citep{gvp+98,ldg+98,ssbe98,gvv+98}. On temporal and spatial grounds,
the physical association of the GRB with the young supernova was
initially controversial \citep{gvv+98,pad+98}, but after a careful
reanalysis of the X-ray data \citep{paa+00} the association of the GRB
with the supernova was confirmed: consistent with the location of the
SN was a slowly variable X-ray source. This transient X-ray source
provided an improved spatial and temporal connection between the GRB
and the SN. It is now widely accepted that GRB\,980425 was coincident
SN 1998bw \citep{kwp+04}.

\begin{figure} 
\centerline{\psfig{file=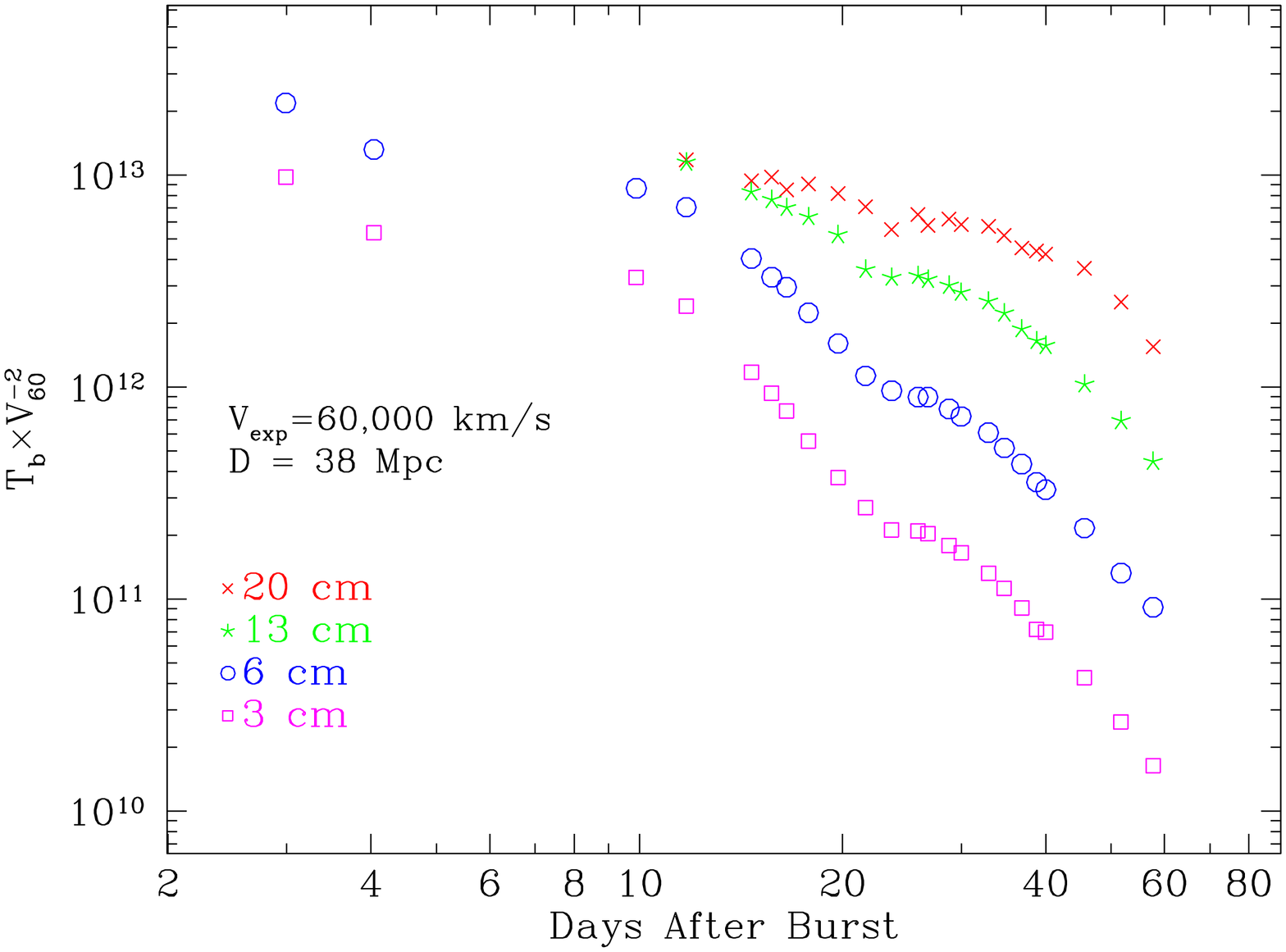,width=4in,angle=270}}
\caption[RSN\,1998bw]{The evolution of the brightness temperature of
radio SN 1998bw, the most luminous RSN ever recorded. The
brightness temperature is computed under the assumption that the radio
photosphere expanded with the same velocity inferred from optical
spectroscopy ($\sim 0.2$c). In order for the true brightness temperature
to be less than the ``Compton Catastrophe'' value ($T_{\rm CC} \approx
10^{12}$ K), relativistic motion in the first week after 980425 is
required. From \citet{kfw+98}} \label{fig:1998bw_radio} 
\end{figure}

The evolution of SN 1998bw was unusual at all wavelengths. The discovery
of prompt radio emission just a few days after the GRB
\citep{kfw+98} (Figure \ref{fig:1998bw_radio}) was
novel. Almost irrespective of modeling assumptions, the rapid rise of
radio emission from SN\,1998bw showed that the time of the SN explosion
was the same as the GRB to  about one day. The brightness temperature
several days after the GRB suggested that the radio photosphere moved
relativistically with $\Gamma \age 3$.  Still, the total energy in
relativistic ejecta was small, $\ale 3 \times 10^{50}$ erg
\citep{lc99}, about two orders of magnitude less than in the SN
explosion itself.

The early optical spectrum of the SN stymied astronomers. \citet{ldg+98}
wrote: {\it``The relative intensity of the different regions of the
spectrum is changing from day to day.  The absence of H lines suggests
that the object is not a Type-II supernova; the lack of Si at 615 nm
indicates that it is not a regular Type-Ia supernova.  The nature of
this puzzling object still evades identification...''}. The initial
IAUC-designated Type was that of a Ib \citep{ssbe98} but a
reclassification to a ``peculiar Type Ic'' was suggested
\citep{pp98,fil98} when no He nor Si II $\lambda$6355 was found.
SN\,1998bw peaked in the V-band 16.2 days (rest frame) after the GRB with
$M_V = -19.16 \pm 0.05 + 5 \log h_{71}$\, mag \citep{gvp+98}. Given the
unknown peculiar velocity of the host and the uncertainty in the
extinction along the line of sight, this absolute peak brightness is
uncertain at the 10\% level.

If GRB\,980425 arose from the $z=0.0085$ galaxy, then it must have
been a very underluminous burst. Assuming isotropic emission, the
energy in gamma-rays was $E_\gamma = 8.5 \pm 0.1 \times 10^{47}$ erg,
more than 3 orders of magnitude fainter than the majority of
long-duration GRBs \citep{fks+01,bfk03}. Any collimation would imply
an even smaller energy release in gamma-rays. Still, many held that on
purely phenomenological grounds that GRB\,980425 and SN 1998bw had to
be physically associated. The SN was simply was too unusual not be
connected with the GRB.

\begin{figure}
\centerline{\psfig{file=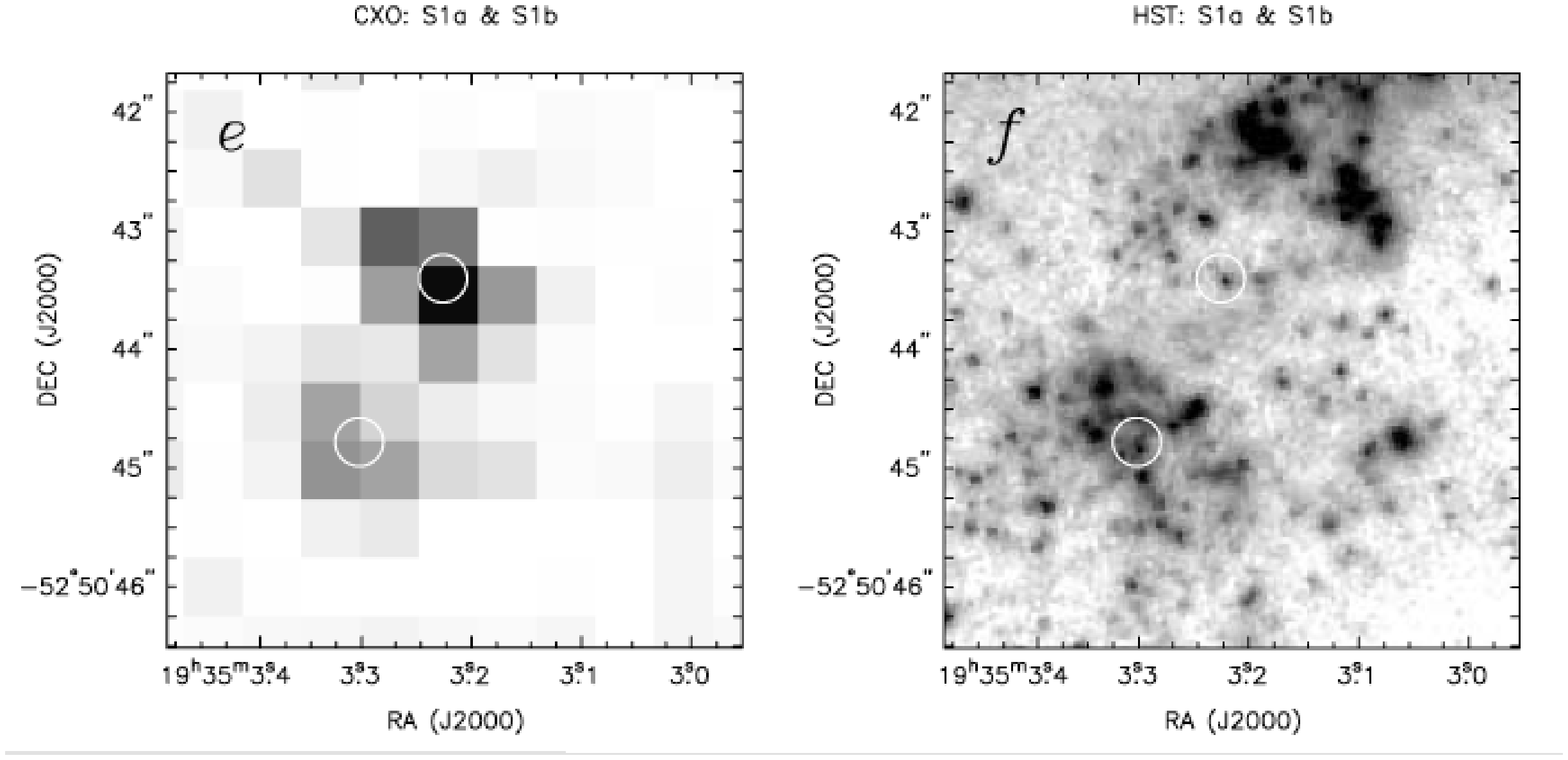,width=5.in} }
\caption[The Home of SN\,1998bw]{The location of SN\,1998bw as viewed at
late times by {\it Chandra} (left) and {\it HST} (right). The circles represent the $1 \sigma$ astrometric position from {\it Chandra}. The fading
X-ray source ("S1a") consistent with SN\, 1998bw is to the south east (bottom left). 
From \citet{kwp+04}} \label{fig:1998bw_home} \end{figure}

Accepting the connection and given the very low redshift, the burst
site of at least one (albeit faint) GRB was studied in unprecedented
detail.  The SN position was found with the {\em Hubble Space
Telescope} (HST) to have been in an apparent star forming region
within 100 pc of several young stars \citep{fha+00}, as expected of
core-collapse SNe. Likewise, late-time Chandra X-ray imaging revealed
an X-ray point source (Figure \ref{fig:1998bw_home}) consistent with
the optical and radio SN position \citep{kwp+04} (and, curiously, a
variable ultra luminous X-ray source in the same spiral arm). Even at
such low redshifts, the late-time HST imaging was incapable of
resolving any star cluster or companion. This fact serves as a bleak
reminder that GRBs which occur at even higher redshifts may not have their
progenitors nor immediate progenitor environments directly observed.

The unusual properties of the event led some to suggest that GRB\,
980425 represented not only a phenomenological sub-class of GRBs, but
one physically distinct from other long-duration GRBs
\citep{bkf+98,nbw99,Woo99,Mat99,Tan01}.  The consensus is now that
both GRB\, 980425 and SN\, 1998bw, in particular the energetics of
such, represent the extreme in a continuum of events all with same
underlying physical model. Indeed GRB\,031203 and its associated SN
(\S \ref{sec:spect}) are now considered the closest cosmological
instance of GRB\, 980425 and SN\, 1998bw \citep{skb+04}.

\subsubsection{Late-time Bumps}
\lSect{bumps}

\begin{figure}[tp]

\centerline{\psfig{file=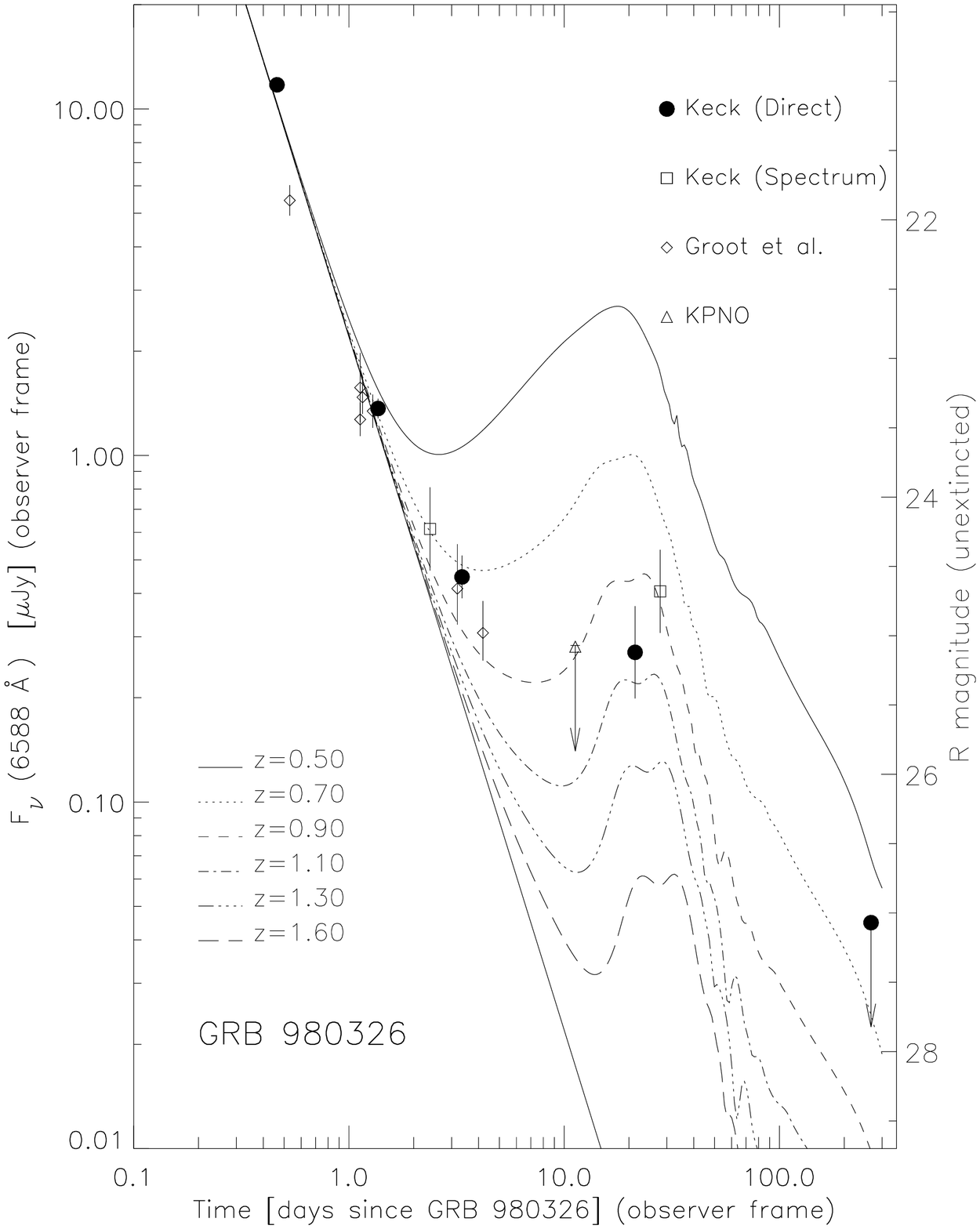,width=4in}}

\caption[GRB 980326]{An optical SN-like bump superimposed on the afterglow of GRB 980326. Models of SN
1998bw at different redshifts are shown. The color and light curve of
the bump was found to be consistent with 1998bw at redshift of unity.
Without a measured redshift, however, there is a degeneracy between the
absolute brightness, distance, and extinction due to dust in the host
galaxy.  From \citet{bkd+99}.} \label{fig:980326}
\end{figure}

Viewing GRB\,980425, and its origin, as distinct from the
``cosmological'' set of GRBs was the norm in 1998. Though the
detection of a contemporaneous SNe would be a natural consequence of a
massive star origin \citepeg{Woo93,han99}, no other GRB had obvious
late-time emission that resembled a supernova. A report of a red
emission ``bump'' following GRB\,980326 \citep{bk98} was interpreted as
due to a coincident SN at about a redshift of unity
\citep{bkd+99,cast99} (Figure \ref{fig:980326}). Without a
spectroscopic redshift for GRB\,980326 and multi-band photometry
around the peak of the bump, the absolute peak brightness and type of
the purported SN could not be known. The available data were also
consistent with a dust echo \citep{eb00,rei01b} or dust re-radiation
\citep{wd00} from material surrounding the GRB. A subsequent
reanalysis of the afterglow of GRB\,970228 revealed evidence for a
bump which appeared to be the same absolute magnitude as SN\,1998bw
with similar rise times \citep{rei99,gtv+00,rlc00}.  Similar reports
of bumps were made
\citep{svb+00,fvh+00,bhj+01,lcg+01,csg+01,sok01,bdf+01,clg+03,gff+05,mpp+05},
but none as significant and with as clear cut connection to SNe as
GRB\,980326 and GRB\,970228 (see \citealt{blo05}).

Concerted multi-epoch ground-based and space-based observing campaigns
following several GRBs strengthened the notion that late-time bumps
were indeed SNe
\citep{pkb+03,bkp+02,gsw+03,sgn+05}. The SN of GRB\,011121 showed a spectral
rollover during peak at around 4000 \AA, nominally expected of
core-collapse SNe in the photospheric phase. The typing of the SN
associated with GRB\,011211 was controversial, with \citet{gsw+03}
showing evidence that the brightness and color evolution resembled
1998S (a Type IIn; see also \citealt{mr04}) and \citet{bkp+02} showing
consistency with a Ic-like curve interpolated between the faint and
fast 1994I and the bright and slow 1998bw. We discuss the overall
census of GRB-SNe photometry in \Sect{allsn}.

\subsubsection{Spectroscopic Evidence and a Clear Case}
\label{sec:spect}

\begin{figure}[tp] \psfig{file=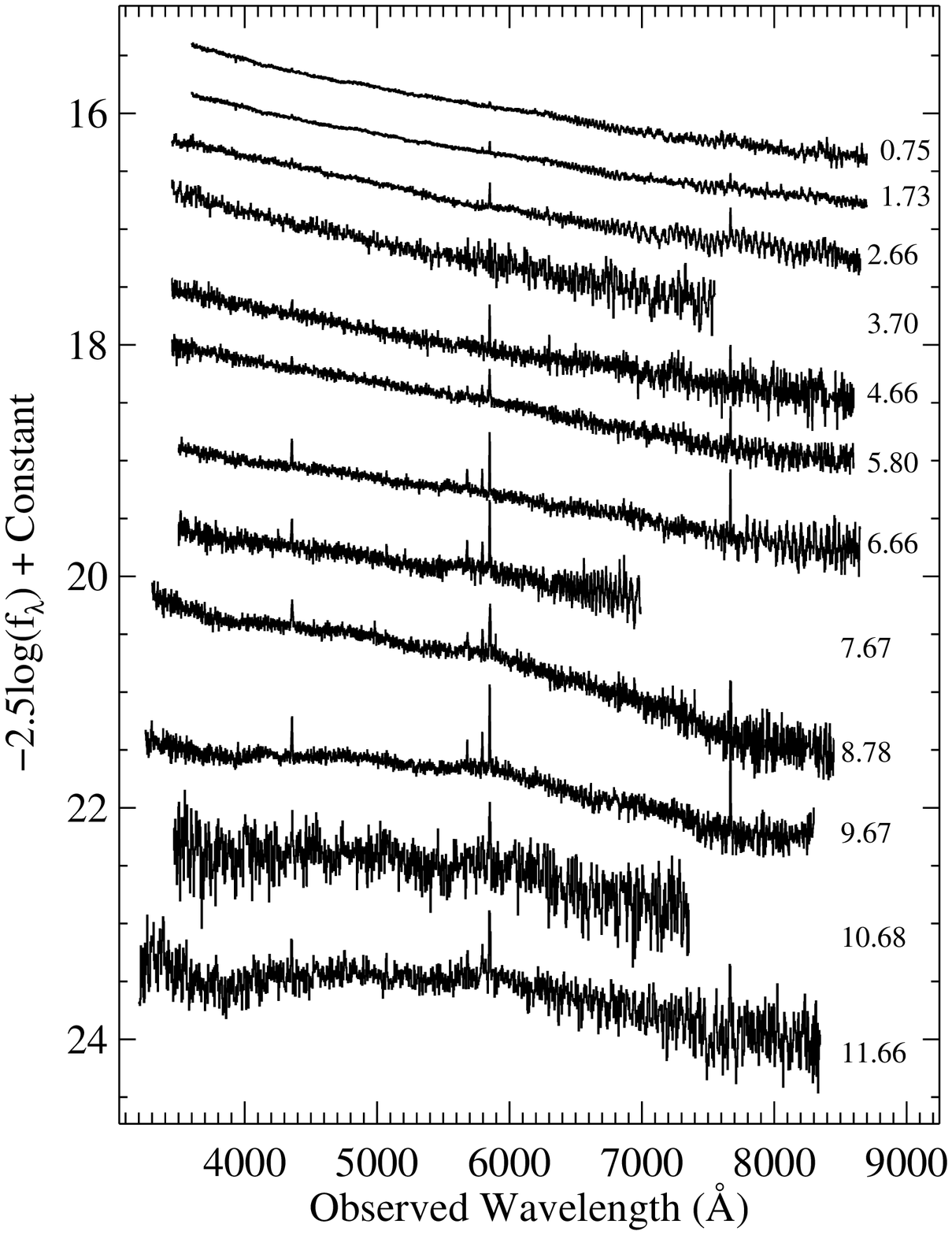,width=4in}
\caption[2003dh emerges from the afterglow of GRB 030329]{The discovery
spectra of the emergence of SN 2003dh from the glare of the afterglow
of GRB\,030329. Shown is the observed spectra, a combination of
afterglow and supernova. Days since the GRB are noted at right. The
narrow emission lines are from the host galaxy and do not change in
intensity throughout. From \citet{math04}.}
\label{fig:2003dh-emerge} \end{figure}

Though several bumps were found with characteristics remarkably
similar to \hbox{Type I} SNe, the first truly solid evidence for a connection
between ordinary GRBs and SNe came with the detection of the
low-redshift ($z=0.1685$; \citealt{gpe+03}) GRB\,030329 and its
associated supernova, SN 2003dh (for recent reviews, see
\citealt{math04} and \citealt{dv05}).  Shortly after its discovery 
(the brightest burst HETE-2 ever saw), the afterglow of the GRB
\citep{pp03,tori03} was very bright ($R \sim 13$ mag). It faded
slowly, undergoing several major re-brightening events in the first few
days \citep{bsp+03,gkr+03,mgs+03,bvb+04,log+04}. Given the low
redshift, several spectroscopic campaigns were initiated.
Spectra of the afterglow 
\citep{mgh+03,gmo+03,mgo+03,cff+03,smg+03,kdw+03,hsm+03}, 
6.6 and 7.7 days after the GRB, showed a deviation from a pure
power-law and the emergence of broad SN spectral features (Figure
\ref{fig:2003dh-emerge}).

\begin{figure}[tp] 
\psfig{file=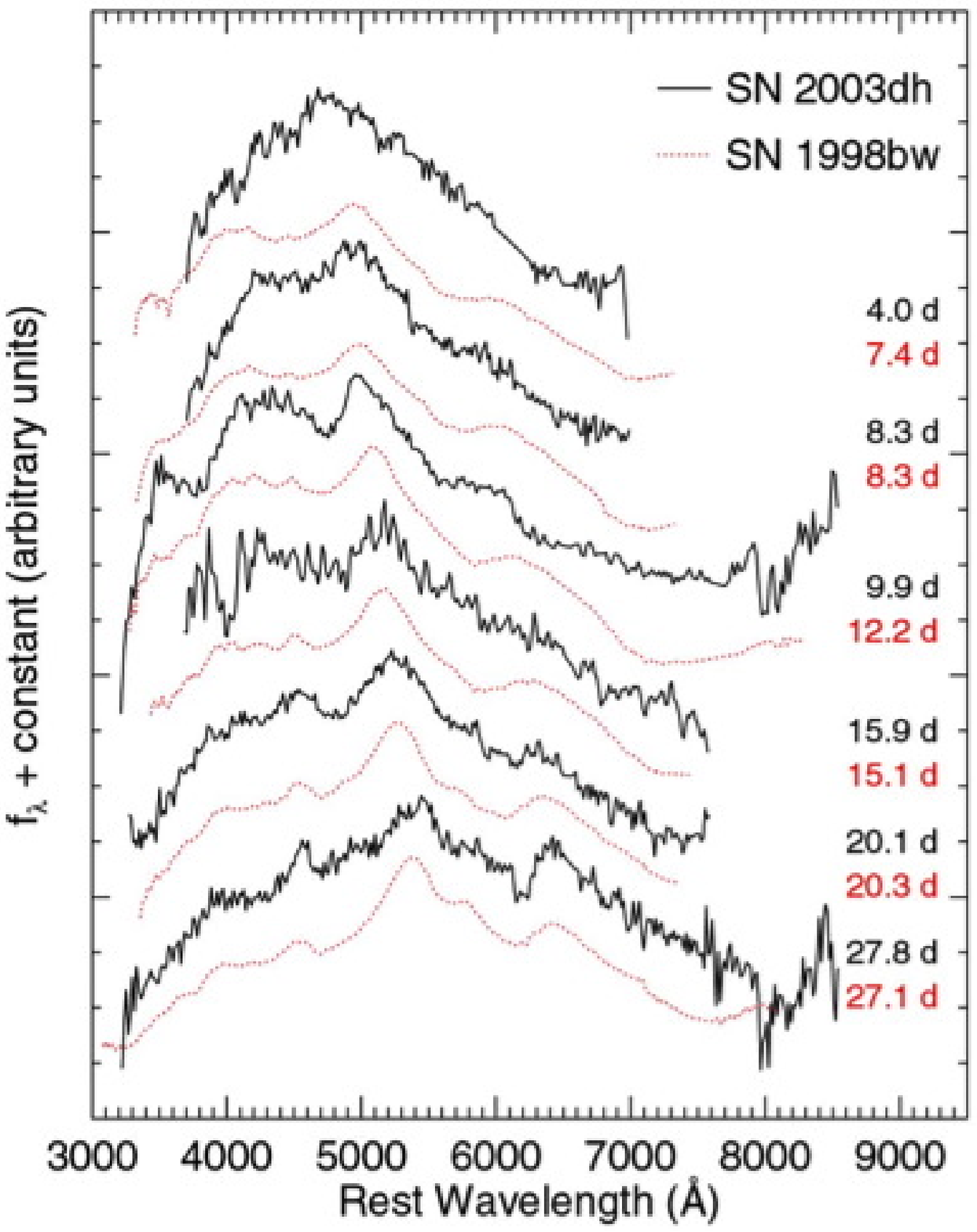} 
\caption[comparing 2003dh with 1998bw]{A comparison of the rest frame 
optical spectrum of SN 1998bw and SN 2003dh (associated with GRB\,
030329).  The spectral features and evolution show a remarkable
similarity suggesting that the SN explosions of cosmological GRBs are
connected to the mechanisms inferred for 1998bw. The afterglow
component has been modeled and removed from the observed spectra. From
\citet{hsm+03}} \label{fig:2003dh-1998bw}
\end{figure}

As the afterglow faded, the SN became more prominent and showed
remarkable similarity to SN 1998bw (Figure \ref{fig:2003dh-1998bw}).
Spectrapolarimetric observations at later times showed that the SN
light was somewhat polarized ($P < 1\%$) indicating mild asymmetry in
the sub-relativistic ejecta. Given the broad spectral features,
indicating high velocities ($\age$ 25000 km s$^{-1}$;
\citealt{smg+03,hsm+03,mdt+03}), and apparent absence of hydrogen,
helium and strong Si II $\lambda$6355 absorption, a classification as
Type Ic-BL was natural \citep{mgs+03}. We leave the discussion of the
modeling of SN\,2003dh to \S \ref{sec:models}, but we
emphasize here that even what should be the easiest-to-measure
``observable'' of a SN-GRB can be highly model dependent: the reported
peak magnitude of SN\,2003dh differed by more than 1 magnitude (from
0.6 mag fainter to 0.5 mag brighter than SN\,1998bw)
\citep{mgs+03,log+04,hsm+03,mdt+03,bvb+04}. In part, the differences
can be due to the quality of the observations near peak, but much of
the difference can be ascribed to different assumptions regarding the
extinction towards SN\, 1998bw, the modeled brightness of the
afterglow at the time of peak, and the modeled $k$-corrections of the
SN (1998bw) template.

There have been a few other reports of spectroscopic identifications
of a SN associated with a cosmological GRB.  A supernova-like
brightening was first reported by \citet{bdb+03} at the position of
the low redshift ($z=0.1055$; \citealt{pbc+04}) GRB 031203.  The
presence of SN 2003lw was confirmed photometrically in multiple
optical and infrared bands
\citep{brf+04,cbd+04,mtc+04,Gal04}. Spectra 17 and 27 days after the
GRB exhibited broad spectral features reminiscent of SN 1998bw at
similar epochs \citep{mtc+04}, but the light curve behavior was more
of a broad plateau around peak than 1998bw
\citep{cbd+04,Gal04,Thom04}. \citet{dmb+03} 
reported both a bump and a low resolution spectrum at the position of
GRB\,021211 ($z=1.006$). A recent claim of a SNe associated with Swift
burst GRB\,050525a has been made based upon a photometric bump and a
low signal-to-noise spectrum near peak \citep{dv06}.

\subsection{SHORT-HARD GAMMA-RAY BURSTS}

\lSect{shorthard}

``Short-hard bursts'' constitute $\sim$30\% of the BATSE sample
\citep{Kou93} and, if, as now appears likely, they typically are sampled 
from a smaller redshift than long soft bursts, they could be the most
frequent form of GRB in the universe.  Some models \citep{Zha03,Wax03}
predict an association of short hard bursts with massive star death,
and hence with SNe.

Recently, however, the counterparts of several short hard bursts have
been discovered \citep{Geh05,bpp+05,Vil05,Fox05,Hjo05,sbk+06}.  These
GRBs have been found at lower redshifts than typical long bursts, but
it is not yet conclusively established that the true bursting rate is
significantly skewed towards lower redshift
\citep{Yam+05,pbc+05,Fox05,bp06}. Short bursts tend to have
prompt burst energy releases much smaller than that of long bursts,
and this may limit their detectability beyond redshifts of unity.
Short bursts have not been found in regions of obvious active star
formation (though two are on the outskirts of a starburst galaxy)
\citep{pbc+05,sbk+06,bp06}. In fact, the hosts of 3 of the 5
well-localized short GRBs are elliptical galaxies (see also
\citealt{bpc+05} and \citealt{bpp+05}). This strongly implicates old
stars or stellar remnants as the progenitors of short-hard GRBs.  This
claim is buttressed by the fact that in at least two cases
(GRB\,050509b: \citealt{bpp+05} and \citealt{hsg+05}; GRB\,050709:
\citealt{Fox05}) the limits on any accompanying supernova are very
tight, stronger than any limits placed on long-duration GRB
counterparts. No supernova is present that is more than 1\% as bright
as SN\, 1998bw ($M_R > -12$ at 16 days in GRB\, 050709;
\citealt{hsg+05}).  Still, the internal-external shock model for
short-duration GRBs appears to accommodate the data
\citep{lrrg05,bpp+05,Fox05,pana06}.

With such a small sample, it would be prudent to wait \citepeg{bp06} before claiming
that all short hard bursts are the result of merging compact objects,
but the data so far are certainly consistent with that hypothesis.  This raises the interesting possibility that we may
need to define yet another class of ``peculiar supernova''
(mini-supernova?), if the radioactive ejecta of the mergers prove
capable of powering a brief optical and X-ray display
\citep{Li98,kul06}.

If short hard bursts are merging compact objects, the duration of the
bursts also has some interesting implications for GRBs in
general. Associating the event duration with the operation of the
central engine implies that the viscous lifetime of any accretion disk
created in the merger, typically 0.1 \Msun \citep{Ros03,Set04} is
$\sim$0.1 s. This timescale is far too short for disk accretion to be
the power source of long soft GRBs unless there is an additional
mechanism for mass accretion onto the disk over longer timescales
(\Sect{models}).

\subsection{COSMIC X-RAY FLASHES}

\lSect{xrf}

Cosmic X-ray flashes \citep[XRFs;][]{hzk+01} are observationally
similar to classic GRBs, only softer, with a similar distribution of
durations \citep{slg+04}. An intermediary in the spectral continuum
between XRFs and classic GRBs are the so-called X-ray rich GRBs
(XRR). Although the cosmological distance scale was well established
\citep{bfvd+03}, it was not until \citet{skb+04a} determined a
spectroscopic redshift ($z=0.251$; XRF\,020903) that the energetics of
any XRF was firmly determined. While the brightness of XRFs implies a
similar energy release (per solid angle) to GRBs, the
internal-external shock model for the prompt and afterglow emission of
XRFs is not as well established. No broadband study of the light curve
of an XRF afterglow has been carried out, so the synchrotron origin,
while consistent with the data, is uncertain.

Because of their similar characteristics to long--soft GRBs, it is
generally thought that the underlying cause is the same
(\Sect{models}), i.e., the explosive death of a massive star. The
emission could be softer because one is just outside the edge of an
ordinary GRB jet \citep{Yam03,Gra05}; because an ordinary GRB jet had
a cocoon of relativistic matter directed towards us with a moderate
Lorentz factor \citep{Zha04}; or because the jet itself had a larger
baryonic loading and hence lower Lorentz factor \citep{Zha04}. The
latter two explanations implicitly assume that the XRF is produced by
an external shock where lower Lorentz factor correlates with softer
spectra. The opposite behavior is expected in the internal shock
model.

An important clue to the origin of XRFs came with the discovery of a
supernova-like bump associated with XRF\,020903 
\citep{skb+04a}. There was a clear rise and decay and, when a low S/N
spectrum was obtained near peak, the galaxy-subtracted spectrum was a
reasonably good match to the spectrum of SN\,1998bw at a similar epoch.
This suggests that at least one XRF originates from the death of a
massive star.

However, aside from XRF\,020903, a concerted search for SN
signatures in XRFs 011030, 020427, 030723, 040701, 040812, and 040916
\citep{Sod2005,lpk+05} turned up no clear evidence for associated
SNe.  A bump peaking in the $R$-band $\sim$16 days after the
XRF\,030723 has been interpreted as a supernova at redshift $\sim$0.5
\citep{fsh+04,tdm+04}. However, the optical spectrum showed no clear
evidence for features and, more importantly, both the $K$-band
(Soderberg, private communication) and $X$-ray light curve
\citep{bss+05} appeared to track the $R$-band light curve. This is
contrary to the expectation from a supernova, where the IR--optical
colors evolve and the IR light luminosity peaks after the optical
light. In other XRFs, no bump was seen. Most constraining is that any supernova in
XRF\,040701 ($z = 0.21$) would have to have been over 3 mag fainter than SN 1998bw
\citep{Sod2005}, fainter than all GRB-SNe known to date\footnote{Note
that no optical afterglow of any sort was seen in XRF\,040701 so the
optical extinction could not be measured as with GRB\, 011121
\citep{pbr+02}.  Therefore this quoted limit for an XRF-SN relies
upon a rather uncertain estimate of the optical extinction based on the
X-ray spectrum. }.  Though fewer bump searches have been conducted for
XRFs than for GRBs, the non-detections are significant because of the
low average redshift of XRFs. The absolute magnitudes probed by deep
(mostly HST) imaging rival all the bump searches in GRBs. Whereas all
GRBs less than redshift 0.7 have {\sl claimed} bump detections, six XRFs
(one with redshift and two more with inferred redshifts less than unity)
show no evidence for a SN-like bump. The search for SNe from XRR GRBs has been more successful, with at least two (041006 and 040924) showing strong evidence for late-time bumps \citep{skp+06}. Both appear at peak to have been fainter than SN\,1998bw.

It may be that the SNe in XRFs are inherently faint (or
absent), which would have important implications for the models, but
the numbers are still small.  Was XRF\,020903 truly an XRF or an
outlier in the classic GRB population? Could the optical extinction for
XRF\,040701 have been greater than estimated? Could the XRFs with no
supernova bump and no well determined redshift be farther away than we
think? The study of XRF related SNe will be a subject of great
interest in the coming years.

\subsection{CHARACTERISTICS OF SUPERNOVAE ASSOCIATED WITH GRBS}

\lSect{characteristics}

The distinguishing feature of a GRB-SN that sets it apart from all
other SNe is the concentration of significant kinetic energy in {\sl
relativistic} ejecta ($\beta \Gamma \age 2$). This does not
necessarily require that the supernova be bright, or even
exceptionally energetic, though GRB-SNe often are. It also does not
preclude the existence of SNe without GRBs, powered by the same energy
source (\Sect{orphan}). But to produce a GRB, one needs at least as
much energy in relativistic ejecta as is observed in $\gamma$-rays and
afterglow emission. That is, $E_{\rm Rel} \gt \, E_{\gamma}$. The
value of $E_{\gamma}$ is difficult to measure directly because of the
effects of beaming, but in typical bursts, it is around $10^{51}$ erg
\citep{fks+01,bfk03}. $E_{\rm Rel}$ can be inferred from radio 
observations at such late times that beaming is no longer important,
and is $\sim 5 \times 10^{51}$ erg \citep{bkp+03,bkf04}. Of course,
there can be considerable variation of both these numbers.

The SNe accompanying GRBs also differ from common SNe
\citep{fil97} in other ways (Table \ref{tab:snprop}), 
most obviously the absence of hydrogen in their spectra: 
GRB-SNe appear to be Type I SNe. Indeed, where spectra of sufficient
quality exist to be sure, the supernova is of Type Ic-BL. Some of the
broad peaks seen in the GRB-SNe spectra are likely due to low opacity,
rather than due to emission from a single ion spread over large
velocity ranges \citepeg{inm+03}. Near maximum light GRB-SNe do appear
to show broad absorption lines of \ion{O}{I},
\ion{Ca}{II} and \ion{Fe}{II} \citep{imn+98}. About 7 days before
maximum, the width of a weak \ion{Si}{II} line in SN 1998bw suggested
expansion speeds in excess of 30,000 km s$^{-1}$
\citep{pcd+01}. There has never been a photospheric spectrum of a
confirmed GRB-SN that indicated the presence of H and no optical
\ion{He}{I} lines have been seen (e.g., $\lambda$6678, $\lambda$7065,
and $\lambda$7281), leading to a classification as a Type
Ic\footnote{An infrared feature observed in SN 1998bw may have been
due to He \citep{pcd+01} but that not a secure identification and,
more important, the classification distinction between Ib and Ic
depends on the observed presence or absence, respectively, of
\ion{He}{I} in the optical waveband (T.\ Matheson, private
communication).}. The late time nebular phases (at least in the case
of SN 1998bw) show lines of [\ion{O}{I}], \ion{Ca}{II}, \ion{Mg}{I},
and \ion{Na}{I D} \citep{shc+02}.

As detailed in \S \ref{sec:allsn}, the {\sl median} peak magnitude of the {\sl observed} GRB-SNe sample is comparable to that of Type Ia SNe. This large apparent brightness could be misleading however, because of the stringent
requirements -- rapidly declining optical afterglow, low red shift,
faint host galaxy -- required for detection (\Sect{allsn}).  In one
case, GRB\,010921, the upper limit on any supernova is absolute
magnitude $-17.7$ and, as noted previously, some of the SNe with XRFs
may be fainter still. Indeed, when the non-detections of GRB-SNe are
accounted for, the inferred ``true'' mean of the sample ($M_V = -18.2
\pm 0.4 + 5 \log h_{71}$) is considerably fainter than the mean of
normal Type Ia SNe.

At late times, the decay of $^{56}$Co often leads to a steady
exponential decline in the light curve of Type Ib SNe, providing all
decay energy remains trapped. SN\,1998bw initially declined somewhat
faster than this, presumably because of $\gamma$-ray escape
\citep{ms99,shc+02}. At very late times ($\age 500$ days), a
flattening seen in the light curve could be interpreted as greater
retention of the energy from radioactive decay as well as the
contribution of species other than $^{56}$Co \citep{shc+02}, but this
could have other explanations.  Other GRB-SNe could not be followed
with sufficient sensitivity to see the exponential tail.

The radio emission of GRB 980425 and SN\,1998bw showed no evidence for
polarization \citep{kfw+98} which suggests that the mildly
relativistic ejecta were not highly asymmetric, at least in
projection. Still, internal Faraday dispersion in the ejecta would
serve to suppress a radio polarization (Soderberg, private
communication). The optical light of SN\,1998bw showed significant
evidence for polarization at the 0.5\% level \citep{pcd+01}, which is
consistent with polarization inferred in other core-collapse SNe
\citep{wwlc96,lfcr02}. This implies some degree of asphericity in the
non-relativistic ejecta \citepeg{hww99}, but, as \citet{pcd+01} noted,
there is a degeneracy between the viewing angle and the level of
asymmetry. Significant polarization was observed for the afterglow of
GRB\,030329 over many epochs, but by the time SN\,2003dh dominated the
optical light, two epochs of observations revealed only marginally
significant polarization \citep{gkr+03} (even then, the afterglow
could have contaminated the polarization signal). While polarization
is surely a critical ingredient towards understanding the SN explosion
geometry, \citet{kpm+04} have pointed out that interstellar dust for
mildly extinguished lines-of-sight should artificially induce
polarization at the 1\% level. Thus, even if polarization is detected
in GRB-SNe, the interpretation is anything but straightforward.

In the case where the brightness is as great as Type Ia, theory
demands a large production of $^{56}$Ni (\Sect{nucleo}).  In order to
make so much $^{56}$Ni, large shock energies and stellar masses are
required, at least in one-dimensional models. The high energy is
consistent with the rapid expansion velocities inferred from the
spectra. The combination -- high mass, large velocity, and big
$^{56}$Ni mass -- implies kinetic energies $E_{\rm SN} \sim 10^{52}$
erg. Since the collapse of the iron core to a neutron star, plus any
accretion into a black hole must release $E_{\nu} \sim 10^{53}$ erg,
one has the interesting energy relation, $E_{\nu} \gg E_{\rm SN} \gt
E_{\rm Rel} \gt E_{\gamma}$. There may be significant gravitational
radiation as well, with $E_{\rm GW}
\sim E_\nu$ \citep{Putb04}. The energy in $\gamma$-rays is 
only a small fraction, $\sim1$\% of the total energy released in the
explosion.

Table \ref{tab:snmodel} gives the energetics and theoretical masses of
$^{56}$Ni produced in each of the three GRBs with definite
spectroscopic SNe. For those GRBs which exhibit jet breaks and are
thought to be observed nearly pole on, an actual $E_{\gamma}$ can be
inferred. In other cases, E$_{\rm Rel}$ probably provides an upper
limit to E$_{\gamma}$, but E$_{\gamma}$ could be much greater than
E$_{\gamma,{\rm iso}}$ if the event was observed off-axis.  SN 2003lw
had a light curve and spectrum similar to SN 1998bw and its energy and
$^{56}$Ni production are assumed here to be the same \citep{Gal04}.
It is noteworthy that E$_{\rm Rel}$ and E$_{\gamma}$ for all three
bursts in Table \ref{tab:snmodel} are much less than the canonical $5
\times 10^{51}$ and 10$^{51}$ erg, respectively, mentioned above 
for common GRBs.  This reflects, at least in part, the greater
likelihood of discovering a supernova if the optical afterglow is
faint.

So far, no clear correlation has been found between GRB properties, or
even $E_{\rm Rel}$, and the brightness or energy of the supernova,
though the simplest theory would suggest that more energy input by the
central engine would make both more $^{56}$Ni and more relativistic
ejecta.

The demographics of local Ibc SNe compared with the GRB rate can be
used to constrain the frequency with which SNe accompany GRBs (whether
the GRB is detected or otherwise). At optical wavebands, Type Ic-BL
SNe like 1998bw comprise about $\sim$5\% of all Type Ibc SNe, which
is to simply assert the unusual nature of 1998bw-like SNe. Radio
surveys of local Ibc SNe at early times, reveal that no more than 3\%
\citep{bkfs03} harbor SNe with relativistic ejecta like SN\,1998bw. 
At late times, the radio emission from GRBs initially directed off
axis should become observable as $\Gamma$ of the shock slows to
unity. Yet none of the local Ibc SNe studied show evidence for an
off-axis jet with large relativistic energy. Specifically, at
the 90\% confidence level less than 10\% of all Ibc harbor an ordinary
off-axis GRB \citep{snkb06}. Moreover, the class of optical BL SNe
cannot all be related to GRBs at the 84\% confidence level
\citep{snkb06}.

\begin{table}%
\def~{\hphantom{0}}
\caption{Properties of good candidate supernovae associated with  GRBs, X-Ray Flashes, and X-Ray Rich GRBs}\label{tab:snprop}
\begin{tabular}{@{}lccccr@{}}%
\toprule
Name          & $z$    &  Peak                  & $T_{\rm peak}^a$ & SN likeness/ & References \\ 
Burst/SN        &        &  [mag]                 &    [day]       & designation  &      \\
\colrule
GRB 980425/1998bw & 0.0085 & $M_V = -19.16 \pm 0.05$         & 17             & Ic-BL        & b  \\ 
GRB 030329/2003dh & 0.1685  & $M_V = -18.8$ to $-19.6$  & 10 - 13        & Ic-BL        & c  \\
GRB 031203/2003lw & 0.1005 & $M_V =-19.0$ to $-19.7$  & 18 - 25        & Ibc-BL       & d  \\ 
                  &        &                        &                &              &    \\
XRF 020903        & 0.25   &  $M_V = -18.6 \pm 0.5$         & $\sim$15       & Ic-BL        & e  \\  

GRB 011121/2001dk & 0.365  & $M_V  = -18.5$ to $-19.6$  & 12 -- 14        & I (IIn?)     & f  \\ 

GRB 050525a & 0.606  & $M_V  \approx -18.8$  & 12        & I     & g  \\ 

GRB 021211/2002lt & 1.00   & $M_U = -18.4$ to $-19.2$  & $\sim$14       & Ic           & h  \\
GRB 970228        & 0.695  & $M_V \sim -19.2$       & $\sim$17       & I            & i  \\
XRR 041006        & 0.716  & $M_V = -18.8$ to $-19.5$       & $16 - 20$           & I            
& j  \\

XRR 040924 & 0.859  & $M_V  = -17.6$   & $\sim$11        & ?    & k  \\ 

GRB 020405        & 0.695  & $M_V \sim -18.7$       & $\sim$17       & I            & l  \\
\botrule
\end{tabular}
$^a$The time of peak brightness is reported in the rest frame if the redshift is known, observed frame otherwise.
$^b$\citet{gvp+98}. 
$^c$\citet{hsm+03,smg+03,bvb+04,log+04}.
$^d$\citet{mtc+04,cbd+04,Thom04,Gal04}.
$^e$\citet{Sod2005}.
$^f$\citet{bkp+02,gsw+03,gks+03}.
$^g$\citet{dv06}.
$^h$\citet{dmb+04}.
$^i$\citet{gtv+00,rei99}.
$^j$\citet{sgn+05,skp+06}.
$^k$\citet{skp+06}.
$^l$\citet{pkb+03}.
\end{table}

\begin{table}%
\def~{\hphantom{0}}
\caption{Physical Properties of GRB-SNe}\label{tab:snmodel}
\begin{tabular}{@{}lcccccc@{}}%
\toprule
GRB/SN        & E$_{\rm SN}$    & E$_{\rm Rel}$    & E$_{{\rm iso}}(\gamma)$& E$_{\gamma}$    & M($^{56}$Ni) & Refs.  \\
              & (10$^{52}$ erg) &  (10$^{49}$ erg) &    (10$^{49}$ erg)    & (10$^{49}$ erg) & ($M_\odot$)  &        \\
\colrule
              &  &  &  &  &  &  \\
980425/1998bw &      2 -- 3$^a$  &     1 $-$ 30       &    0.06 -- 0.08      &   $<$0.08       &   0.5 -- 0.7  & b  \\ 
030329/2003dh &      2 -- 5      &     $\approx$50    &       1070           &     7 -- 46      &   0.3 -- 0.55 & c  \\
031203/2003lw &      2 -- 3      &      2             &    $2.94 \pm 0.11$   &   $<$3       &   0.5 -- 0.7  & d  \\ 
              &  &  &  &  &  &  \\
\botrule
\end{tabular}
$^a$\citet{hww99} inferred a kinetic energy for SN 1998bw one order of 
magnitude smaller based upon highly asymmetric model.

$^b$\citet{imn+98,Woo99,lc99,gvv+98,fb05}

$^c$\citet{mdt+03,bkp+03,Woo05,Den05,fsk+05,fb05}

$^d$\citet{Mal04,Gal04,log+04,skb+04,sls04,fb05}

$^e$The absence of increasing energy inferred in the afterglow
blast wave at late times, suggests that these sources were not off-axis
GRBs \citepeg{skb+04} and therefore that $E_{\rm iso}(\gamma)$ is indeed
an upper limit to the true energy released in $\gamma$-rays.

\end{table}

\subsection{DO ALL GRBs HAVE SUPERNOVA COUNTERPARTS?}
\lSect{allsn}

While 2003 was a banner year for the SN-GRB connection, there has
since been a noticeable lack of new GRBs with spectroscopic
SNe. Indeed, {\sl most} long-duration GRBs do {\sl not} have a
detected associated supernova. Yet the indirect evidence
(\Sect{hosts}) supports the consensus view that most long-duration
GRBs arise from the death of massive stars. These two statements can
be reconciled by invoking observational biases which are either
extrinsic or intrinsic to the explosions.

\begin{figure}[tp] \psfig{file=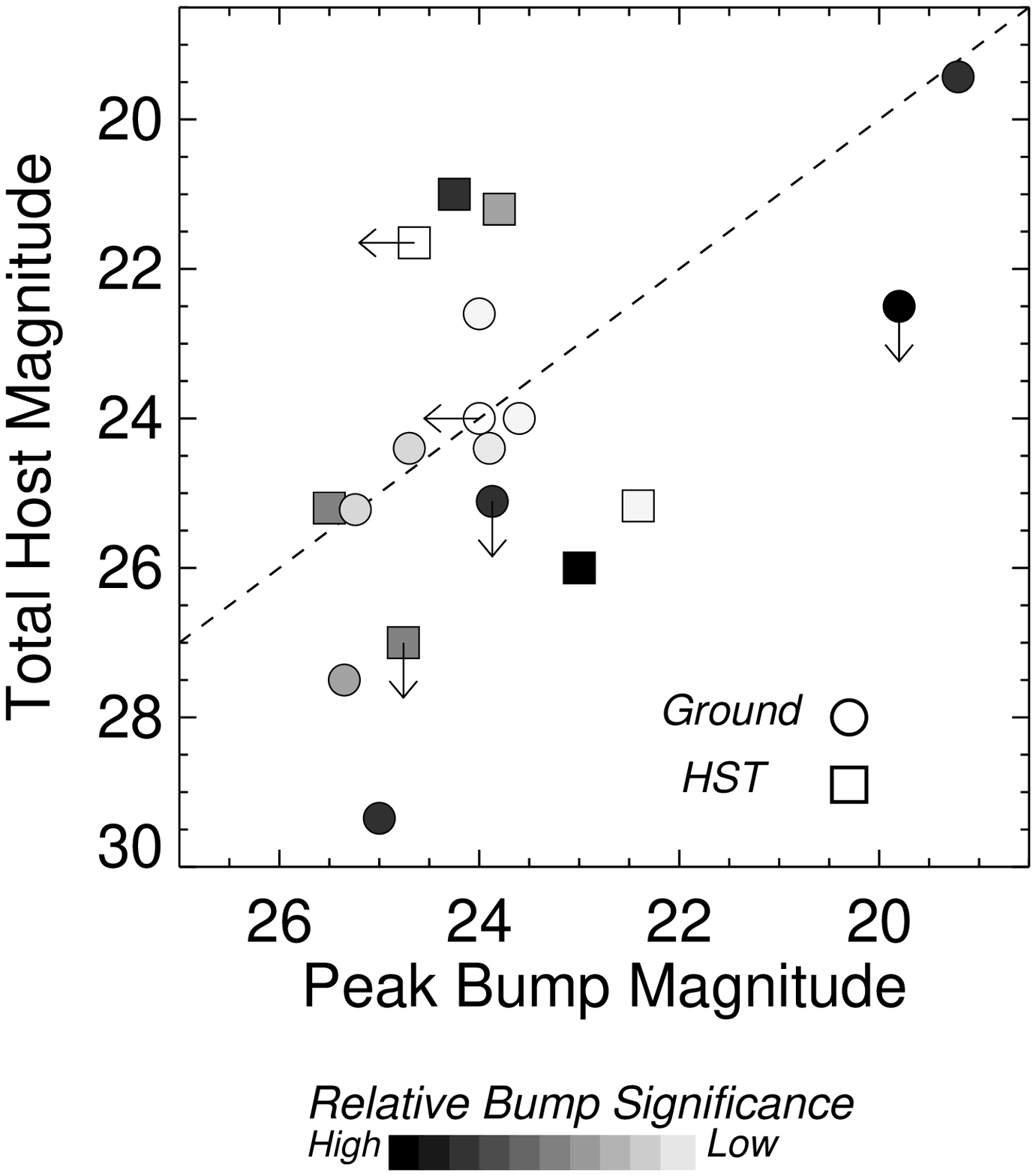,width=4in}
\caption[Bumps versus hosts]{Illustration of the difficulty in finding
GRB-SNe: claimed bump peak magnitudes versus the integrated magnitude
of the host galaxy. Relative significances of the detections are based
on our subjective analysis of the believability of the bumps. Many
bumps were claimed by differencing catalog magnitudes --- small
systematic errors in measuring the true host magnitudes artificially
reveal bumps.  High resolution imaging ameliorates some of the endemic
problems of bump detections from catalogs.} \label{fig:snhost}
\end{figure}

Extrinsic biases, those that result in the decreased probability for
discovery even if the supernova is bright, are straightforward to
enumerate:

\begin{enumerate}

\item {\it Localization -} Poor localizations of bursts from their 
afterglows dramatically hamper the ability for large aperture
telescopes to discover emerging SNe. These poor localizations can be
endemic to the detection scheme (e.g., BATSE) or because bursts are
found in regions of the sky that are unfavorable for optical or X-ray
followup (e.g., near the Sun). Afterglows are less likely to be
found near full moon because of the brighter sky and less sensitive IR
detectors. Likewise, if the bump peaks near full moon then the point
source sensitivity is reduced.

\item {\it Dust -} Fainter SNe are expected from bursts
that occur near the line-of-sight through the Galaxy, or in especially
extinction-riddled regions of their hosts. GRB\,021211 and GRB\,031203
were behind significant Galactic columns, which diminished the
sensitivity of the supernova observations.

\item {\it Redshift and luminosity distance -} With increasing luminosity
distance comes higher distance modulus, but the $k$-corrections at
optical wavelengths are particularly unkind beyond redshifts of $z
\approx 0.5$. Absorption line blanketing due to metals (e.g., Fe)
suppress the emissivity below the blackbody luminosity bluewards of
$\sim$4000\AA\ making Type Ibc SNe especially difficult to detect at
increasing redshift \citep{bkd+99}. We expect essentially no optical
flux from GRB SNe at $z \age 1.2$. Since most Swift bursts have been
found above redshift of unity, it is disappointing, but not
surprising, that only one SN (associated with GRB\,050525a) was found
in the first 14 months of Swift observations.

\item {\it Host galaxies -} The host galaxies of GRBs, while generally
  fainter than $L_*$, can still contaminate the light of GRB SNe at
  late times. If all hosts were 0.1 $L_*$, then the integrated light
  of hosts would be $M_V(\rm host) \approx -19$\,mag, comparable to
  the brightest SNe. Thus, without high resolution imaging to resolve
  out diffuse host light, or high signal-to-noise image differencing,
  finding the supernova point source is all but impossible for SNe
  that peak at magnitudes fainter than the integrated light of the hosts.
  Figure \ref{fig:snhost} makes this point rather dramatically. An
  important demonstration is that the SN associated with GRB\,031203
  was only 0.22 magnitudes brighter than the host galaxy at peak. Had
  this burst originated from higher redshift, the photometric
  sensitivity would be diminished, and the SN might not have been
  discovered.

\end{enumerate}

The biases against detection introduced by intrinsic properties of the
explosions are less tractable. Local observations of Type Ibc SNe show
a more than 5 magnitude spread in peak brightness, likely related the
diversity of dust extinction, the spread in explosion energy, and
$^{56}$Ni production. Rise times also range from $\sim$week to several
weeks.  SNe that make the same mass of $^{56}$Ni but which peak later
are fainter.  An intrinsically faint supernova obviously has less of a
chance of being detected. The GRB afterglow brightnesses at late times
may also be comparable to or brighter than the peak of a supernova. It
is again sobering to note that the supernova associated with
GRB\,030329 might never have been recognized at higher redshift. SN
2003dh was discovered spectroscopically when the SN contributed less
than 5\% of the total flux
\citep{mgs+03}. If there is no obvious bump in the light curve
when the supernova peaks \citepeg{log+04}, a photometry-only campaign
could miss the SN altogether.  Still, the supernova might be recovered
with precision color photometry.

\begin{figure}[tp] 
\psfig{file=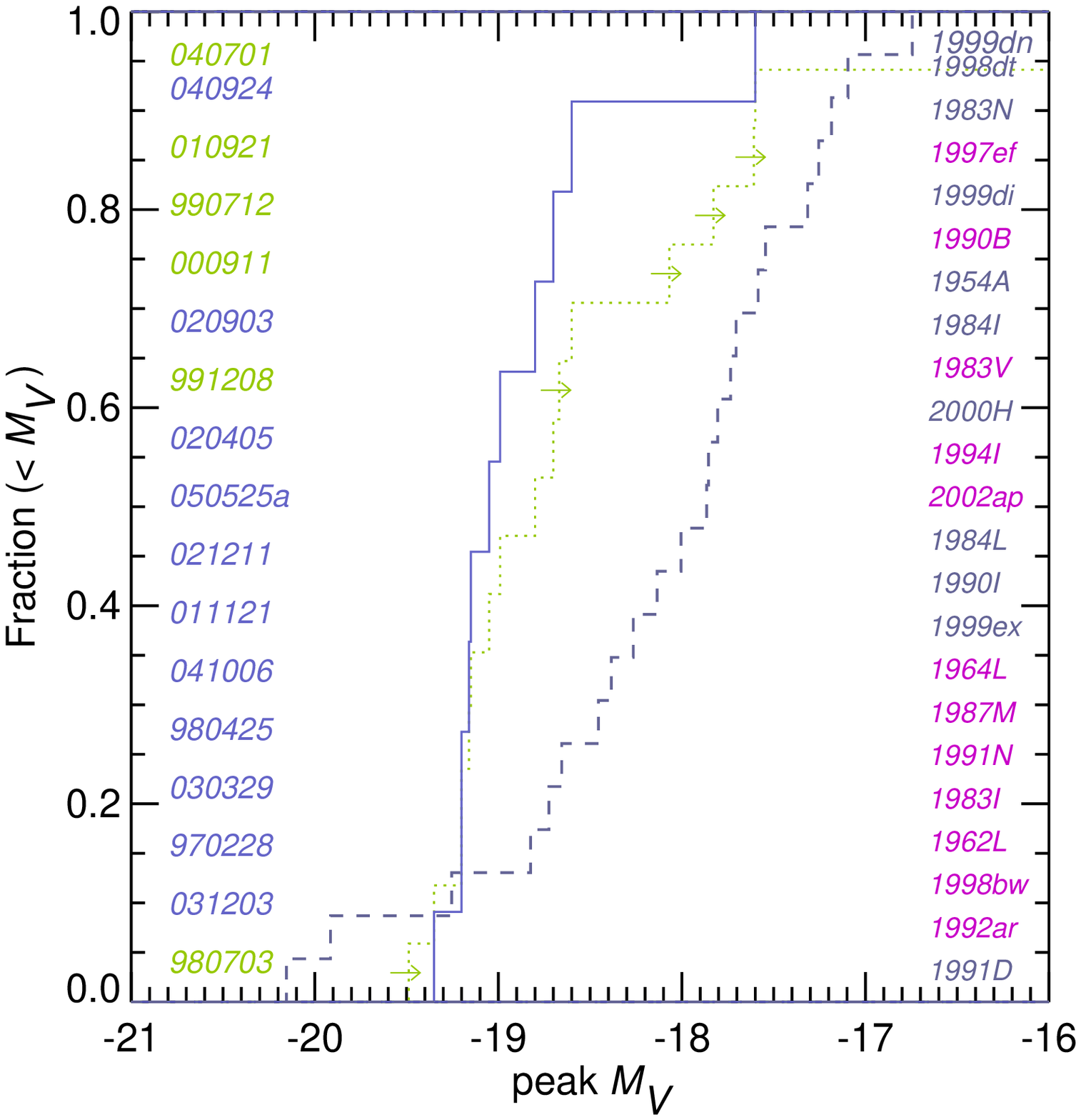,height=4in}

\psfig{file=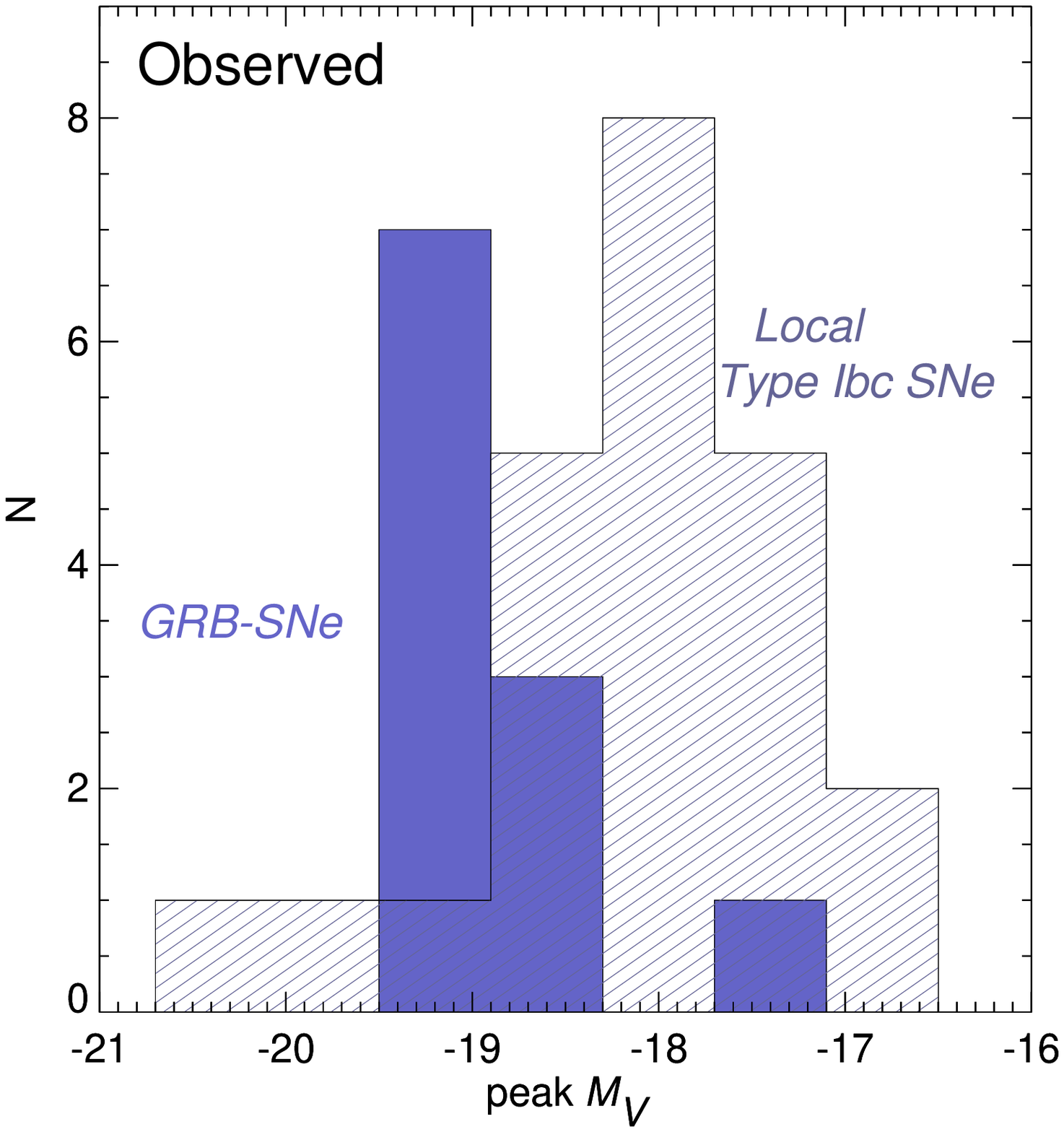,height=4in}

\caption[]{Comparison of the peak $M_V$ of GRB-SNe to the local Ibc population. 
(Top) Cumulative distribution of the Ibc SNe from \citet{rbb06} shown
as a dashed line. The distribution of the {\it observed} GRB-SNe are
shown as a solid line; we include SNe from XRFs and XRR GRBs.  The
dotted line include those GRBs with no significant bump detection or
bumps which do not resemble SNe. The names of the associated GRB are
shown at left and the names of the Ib (grey) and Ic (mauve) are shown
at right. (Bottom) Histogram of the secure bumps as compared to the
local Ibc population. GRB-SNe populate the ``bright''
subclass of the \citet{rbb06} sample, but, as we show in the text, the
bump population is consistent with being drawn from the
entire local Ibc population.} \label{fig:grbsne} \end{figure}

\citet{zkh04} published an important photometric study of
bumps in GRBs, fitting 21 of the best sampled afterglows and finding
evidence for 9 bumps. Statistically significant evidence for bumps are
found in 4 GRB afterglows (990712, 991208, 011121, and 020405) while 5
have marginal significance (970228, 980703, 000911, 010921, and
021211). All of these GRBs had bumps claimed prior to the Zeh
analysis. \citet{zkh04} emphasize that all GRBs with $z \ale 0.7$
appear to have bumps, which of course, would be expected if all
long-duration GRBs have associated SNe. However, bump detection does not
necessarily imply a SN detection. There are, in fact,
important cases where multi-band photometry has shown that late-time
bumps may not be due to a supernova. The late-time light curves of
GRB\,990712 and XRF\,030723, for example (see \Sect{xrf}), do not
appear consistent with a supernova. Figure \ref{fig:grbsne} shows the
results of our compilation of bumps and upper-limits from the
literature.

If GRB-SNe are a particular subset of Type Ibc SNe, then it is useful
to ask if the GRB-SNe sample draws only from the bright end of the
Type Ibc distribution \citep{blo05,skp+06}. \citet{rbb06} report the
detailed modeling of 24 local Type Ibc supernovae and derive the
distribution of peak absolute $M_V$ accounting for Galactic and host
extinction\footnotemark\footnotetext{In what follows, the
\citeauthor{rbb06} magnitudes have been re-calibrated to $H_0 = 71$ km
s$^{-1}$ Mpc$^{-1}$. We have not included the peak $M_V$ from
SN\,1999cq because that value is degenerate with 1994I (T.\ Matheson,
private communication). The sample compiled by
\citet{skp+06} contains a subset (13) Ibc SNe.}. Figure \ref{fig:grbsne},
comparing the observed GRB-SNe population with the Richardson sample,
makes clear that the {\it observed} GRB-SNe are at the bright end of
the {\it observed} Type Ibc population. Here we use the GRB-SNe compilation 
from Table \ref{tab:snprop}; in the case where a range of $M_V$ was reported in the literature, we assume that $M_V$ is the average of the two measurements that define that range (following from 1998bw, we also assume that $M_U - M_V$ = 0.19 mag for GRB\, 021211). The {\sl median} of the
{\sl observed} GRB-SNe sample is $M_V = -19.1 + 5 \log h_{71}$ mag whereas the {\sl median} peak magnitude in the local Ibc sample is $M_V = -17.9 + 5 \log
h_{71}$ mag.

A more subtle but important question is whether the {\it true} GRB-SNe
distribution in peak magnitude, when non-detections are folded in, are
consistent within having been drawn from the local Type Ibc
population. Treating lower limits and insignificant bump measurements
from the literature as non-detections of GRB-SNe, we find that the
{\sl mean} of the true underlying peak magnitude distribution (from the
Kaplan-Meier estimator) is $M_V = -18.2 \pm 0.4 + 5 \log h_{71}$ whereas the
{\sl mean} peak magnitude in the local Ibc sample is $M_V = -18.3 \pm 0.2 +
5 \log h_{71}$\, mag. By all relevant ``survival analysis'' tests
\citep{lif92}, we conclude that the GRB-SNe population is
statistically consistent with having been drawn from the same
population as the local Ibc from \citeauthor{rbb06}\footnote{That is
the probability that the observed deviation is due to random chance is
greater than 0.3 for all tests (e.g., the Generalized Wilcoxon Tests).
We have assumed no censoring of the local Ibc data; since bright Ibc
systematically detected over faint Ibc, this clearly biases the Ibc
sample to brigher magnitudes.}. Moreover, although the GRB-SNe
population is consistent with both the Ib and Ic sample, the
connection with Ic SNe is more heavily favored\footnote{Specifically,
$P$(Gehan's Generalized Wilcoxon Test) = 0.14 comparing the GRB-SNe
sample to local Ib SNe and $P$(Gehan's Generalized Wilcoxon Test) =
0.96 when comparing to local Ic SNe. Other tests show similar
improvements for the Ic comparison.}.

\section{MODELS}

\lSect{models}

\subsection{CORE-COLLAPSE SUPERNOVA MODELS}
\lSect{corecollapse}

Models for ordinary core-collapse supernovae have been extensively
reviewed by
\citet{Woo86,Bet90,Bur00,Woo02,Bur03,Bura05,Woo05b,Jan05,Mez05}.  The
current ``standard model'', by no means universally accepted
\citep{Whe05}, begins with the collapse of the iron core of a highly
evolved star that had a main sequence mass of over 10 \Msun. The
collapse, triggered by electron capture and the partial
photodisintegration of the iron at temperatures T$\sim 10^{10}$ K and
densities $\rho \sim 10^{10}$ g cm$^{-3}$, continues until the center
of the central core exceeds nuclear density by a factor of about
two. The rebound, generated by this overshoot and the short range
repulsive component of the nuclear force, launches a shock wave, but
this ``prompt'' shock wave quickly loses all outward velocity to due
to photodisintegration and neutrino losses.  By $\sim$0.1 sec after
the onset of the collapse, one has a ``proto-neutron star'' with
radius $\sim 30$ km and mass 1.4
\Msun \ with a standing accretion shock at $\sim 150$ km through which
matter is falling at about 0.1 - 0.3 \Msun \ s$^{-1}$. 

Over the next tenth of a second or so, the neutron star radiates a
small fraction of its binding energy as neutrinos, $L_{\nu} \sim
10^{53}$ erg s$^{-1}$. Approximately 10\% of these capture on nucleons
in the region between the neutron star and accretion shock.  This
energy deposition drives vigorous convection which helps transport
energy to the shock and also keeps the absorbing region cool enough
that it does not efficiently re-radiate the neutrino energy it
absorbed.  If $\sim$10$^{51}$ erg can be deposited in a few tenths of
a second, the accretion can be shut off. The continuing neutrino
energy deposition then inflates a bubble of radiation and pairs that
pushes off the rest of the star making the supernova. If not,
accretion continues until a black hole is formed. In this standard
model, rotation and magnetic fields are assumed to have negligible
effect.

The problem with this scenario, as many have noted, is that it not
robust in the computer simulations. More often than not, the neutrino
energy deposition, by itself, fails to launch and sustain an outbound
shock of greater than about 10$^{51}$ erg, as is required by analysis
of SN 1987A \citep[e.g.][]{Bet90}, observations of the light curves of
ordinary Type IIp supernovae \citep{Woo86,Elm03,Chi03}, supernova
remnants and ISM heating \citep{Dic93,Tho98}, nucleosynthesis
constraints \citep{Woo95}, and neutron star masses
\citep{Tim96}. If the explosion energy is too weak, large amounts of
matter fall back producing a black hole and robbing the Galaxy of the
necessary iron and other heavy elements. As of this writing it remains
unclear whether the problem with the models is ''simply'' one of
computational difficulty\footnote{The correct calculation must be
done in three dimensions to capture the complex fluid flow in the
convective region where the neutrinos deposit their energy, and the
neutrino transport itself must be followed in great detail and coupled
to the hydrodynamics} or whether key physics is lacking. If additional
physics is required, rotation and magnetic fields are the leading
candidates. However, it remains possible that the answer is being
affected by uncertainties in the high density equation of state,
changes in fundamental particle physics (especially neutrino flavor
mixing), or the possible role of vibrational energy \citep{Bur06}.

\subsection{THE GRB CENTRAL ENGINE}
\lSect{grbmodels}

The general theory of GRBs, with some discussion of models has been
recently reviewed by \citet{Mes02} and \citet{Pir05}.  Unlike a model
for supernovae alone, a viable GRB model must deliver, far away from
the progenitor star, focused jets with at least 200 times as much
energy in motion and fields as in rest mass. The jet typically must
have an opening angle $\sim 0.1$ radian and a power $\sim 10^{50}$ erg
s$^{-1}$. In addition, at least occasionally, the model must deliver
$\sim 10^{52}$ erg of kinetic energy to a much larger solid angle
($\sim$1 radian) to produce supernovae like SN 2003dh and SN
1998bw. This is ten times more than an ordinary supernova.

\begin{figure} 
\centerline{\psfig{file=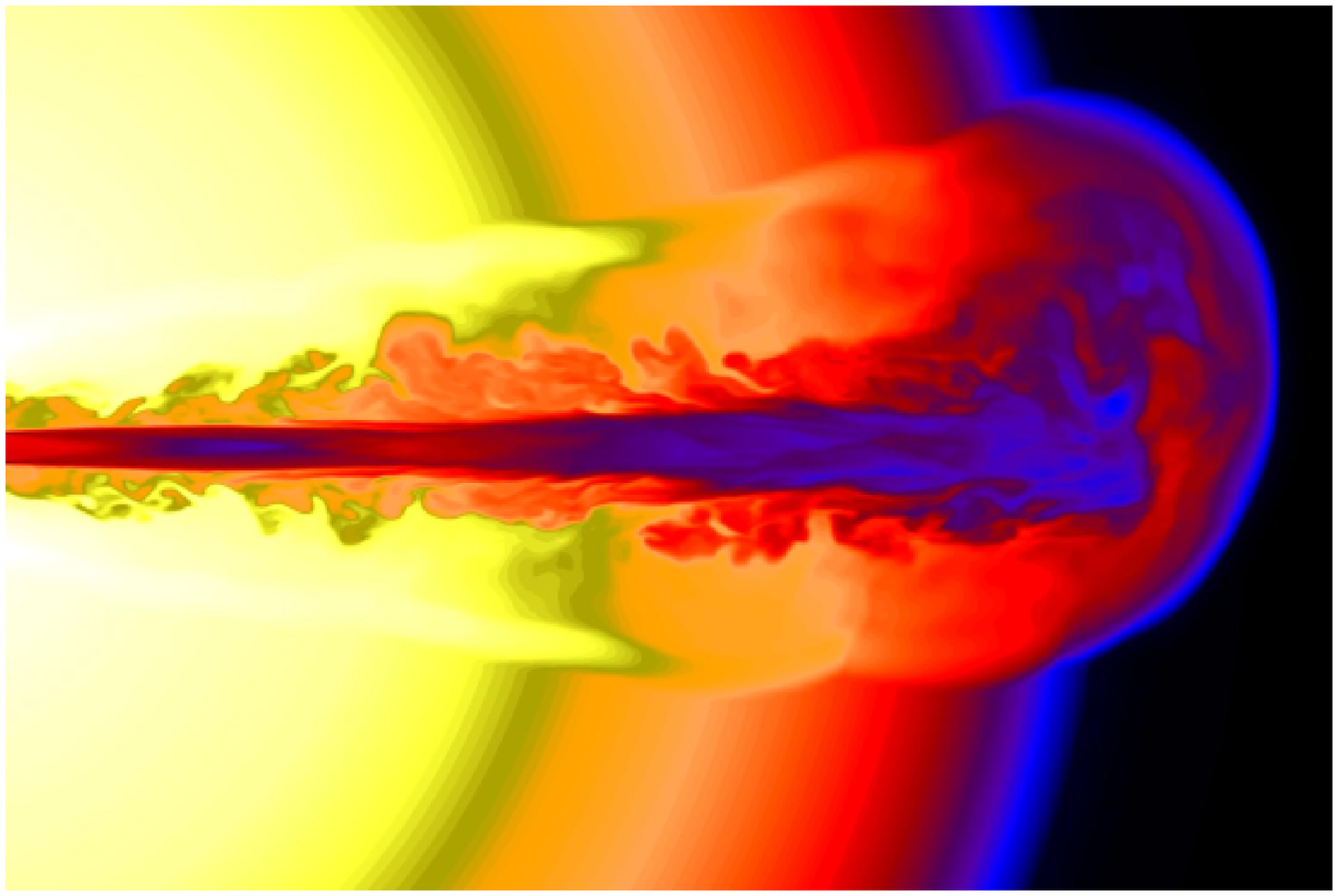,width=4in,angle=0}}
\caption{Break out of a relativistic GRB jet with energy 
$3 \times 10^{50}$ erg s$^{-1}$ 8 seconds after it is launched from
the center of a 15 \Msun \ Wolf-Rayet star. The radius of the star is
$8.9 \times 10^{10}$ cm and the core jet, at infinity, will have a
Lorentz factor $\Gamma \sim 200$. Note the cocoon of mildly
relativistic material that surrounds the jet and expands to larger
angles. Once it has expanded and converted its internal energy this
cocoon material will have Lorentz factor $\Gamma
\sim$ 15 -- 30. A off-axis observer may see a softer display dominated
by this cocoon ejecta. If the star were larger or the jet stayed on a
shorter time, the relativistic core would not emerge, though there
would still be a very energetic, highly asymmetric
explosion. \citep{Zha04}} \label{fig:breakout}
\end{figure}

Except in the {\sl supranova} model (\Sect{supranova}), the fact that a
relativistic jet must escape the host star without losing too much of
its energy severely constrains the model for the central
engine. \citet{Zha04} have shown that the jet head travels
significantly slower than the speed of light and requires 8 -- 25 s to
reach the surface for jet energies from $3 \times 10^{48}$ erg s$^{-1}$ to
$3 \times 10^{50}$ erg s$^{-1}$ (Figure \ref{fig:breakout}). If the jet is
interrupted or changes its orientation significantly in that time, the
flow rapidly degrades to subrelativistic energies and is incapable of
making a GRB. Thus acceptable models must provide $\gtaprx 10^{50}$
erg s$^{-1}$ of relativistic, beamed energy for $\gtaprx 10$ s. If,
for example, one accepts that the duration of short hard bursts ($\sim
0.3$ s) reflects the activity of a central engine, the energy source
for short hard bursts and long soft ones cannot be the
same. Similarly, any model that delivers its energy impulsively, on a
time scale much less than 10 seconds, will not make a GRB, even if
that energy is initially large and highly focused.

\subsubsection{The Millisecond Magnetar Model}
\lSect{msmagnetar}

Building on earlier models of electromagnetic explosions
\citep{Uso92,Tho94,Mes97}, many groups \citep{Whe00,Dre02,Lyu01,Lyu03a} have developed models in which the energy source for GRBs is the
rotation of a highly magnetized neutron star with an initial period of
about one ms (i.e., rotating near breakup). For a rotational velocity
$\Omega \sim 5000$ rad s$^{-1}$ and a dynamo-generated magnetic field,
$B \sim 2 \times 10^{15}$ G, the rotational energy, $E \sim I
\Omega^2/2 \sim 10^{52}$ erg\footnote{The moment of inertia of a neutron star 
for an appropriate range of masses and radii is 0.35 MR$^2$
\citep{Lat01}} and the dipole spin down luminosity, $L \sim B^2 R^6
\Omega^4/c^3 \sim 10^{50}$ erg s$^{-1}$, are typical of GRBs and the
supernovae that accompany them. \citet{Tho04} have considered the
coupling between the neutrino-powered wind that must accompany any
proto-neutron star going through its Kelvin-Helmholtz contraction
\citep{Dun86,Qia96} and the strong magnetic field of a ms
magnetar. Large powers, up to 10$^{52}$ erg s$^{-1}$ can, in
principle, be extracted by a centrifugally driven wind.  The strength
of these models is that they relate GRBs to the birth of an object
known to exist, the magnetar, with an energy scale that is about right
for a neutron star rotating near break up. The Poynting flux models
further offer the possibility of highly energetic outflows with
essentially no limit on the Lorentz factor
\citep[][]{Bla02}. The fields required $\gtaprx 10^{15}$ G are large,
but no larger than in other models.

So far, however, all these models ignore the accretion, $\gtaprx$0.1
\Msun \ s$^{-1}$, that occurs onto the proto-neutron star for several
seconds before it contracts to its final radius and develops its full
rotation rate. This accretion must be reversed before the neutron star
becomes a black hole. The models are also characterized by a
monotonically declining power. Though they do not, so far, consider
the two events separately, an initial blast to a large solid angle is
probably necessary to explode the star and make the necessary
$^{56}$Ni (\Sect{nucleo}). A declining power would be available 10 s
later to make the GRB itself. It may be that the neutron star models
require the initial operation of a successful neutrino-powered
explosion before they can function \citep{Fry04}.

\subsubsection{Collapsars}
\lSect{collapsar}

The necessary conditions to make a collapsar are black hole formation
in the middle of a massive star and sufficient angular momentum to
make a disk around that hole \citep{Woo93,Mac99}. The angular momentum
needed is at least the value of the last stable orbit around a black
hole of several solar masses, $j = 2 \sqrt{3} GM/c = 4.6 \times
10^{16} M_{\rm BH}/3$ cm$^2$ s$^{-1}$ for a non-rotating hole (where
$M_{\rm BH}$ is the black hole mass in solar units) and $j =
2/\sqrt{3} GM/c = 1.5 \times 10^{16} M_{\rm BH}/3$ cm$^2$ s$^{-1}$ for
a Kerr hole with a=1.  This compares with an angular momentum in the
ms magnetar model of $j = R^2 \Omega \sim 5 \times 10^{15}$ cm$^2$
s$^{-1}$ if $\Omega \sim 5000$ rad s$^{-1}$ and R = 10 km. Since the
black holes in the collapsar model are typically very rapidly rotating
and since the specific angular momentum at 3 \Msun \ is always greater
than that at 1.5 \Msun, the minimum angular momentum requirements of
the collapsar and ms magnetar model are similar. Interestingly, there
may also be a {\sl maximum} value of $j$ for the collapsar to work
\citep{Mac99,Nar01,Lee05}. If $j$ is too great, the disk forms at too
great a radius to effectively dissipate its binding energy as neutrino
emission and photodisintegration.  The collapsar model additionally
invokes the formation of a black hole which seems likely above some
critical mass (\Sect{mass}).

\begin{figure} 
\centerline{\psfig{file=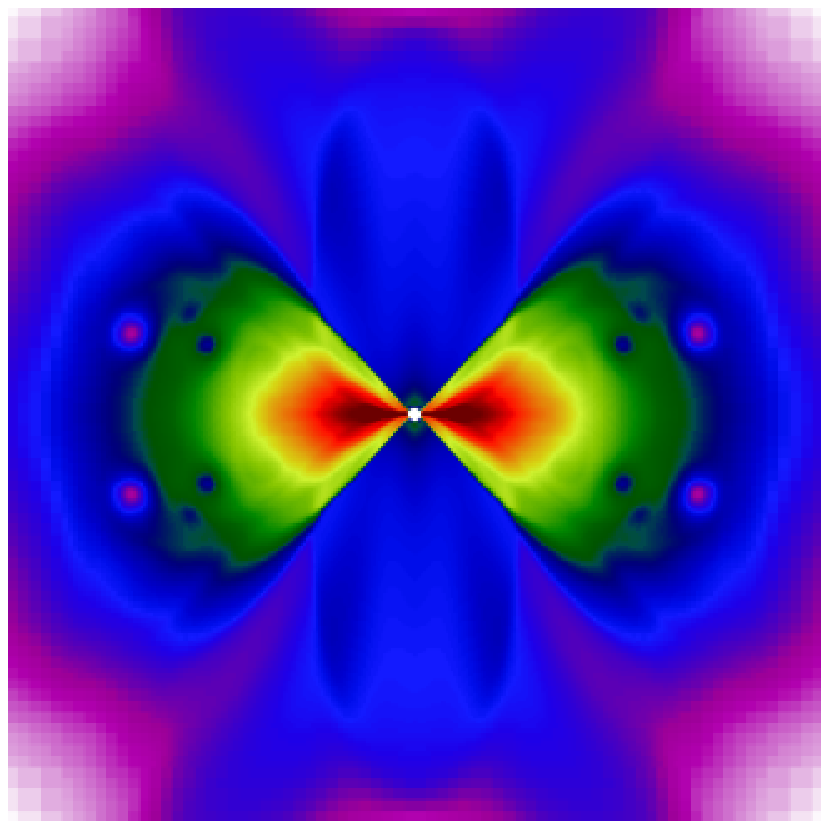,width=4in,angle=0}}
\caption{Collapse of the core of a rapidly rotating 14 \Msun\ Wolf-Rayet star.
Twenty seconds after collapse, a black hole of 4.4 \Msun\ has formed
and has accreted at $\sim$0.1 \Msun\ s$^{-1}$ for the last 15
seconds. The figure is 1800 km across and the inner boundary is at 13
km. Colors indicate density on a logarithmic scale, with the highest
density in the equatorial plane near the black hole being $9 \times
10^8$ g cm$^{-3}$. \citep[][Zhang, Woosley, \& MacFadyen, 
in preparation.]{Mac99}} 
\label{fig:colpsrdisk} 
\end{figure}

Provided a disk and hole form (Figure \ref{fig:colpsrdisk}), the
greatest uncertainty in this model is the mechanism for turning disk
binding energy or black hole rotation energy into directed
relativistic outflows. Three possible mechanisms are discussed: 1)
neutrinos \citep{Woo93,Pop99,Nar01,Dim02}; 2) magnetic instabilities
in the disk \citep{Bla82,Pro03}; and 3) MHD extraction of the
rotational energy of the black hole \citep{Bla77,Lee00,Miz04}. In the
first case, neutrino pairs are generated in the hot disk and impact
one another with the greatest angle along the rotational
axis. Efficient energy deposition is favored by large angle collisions
and the small volume of the region, especially for Kerr black holes,
but probably the efficiency is no greater than $\sim$1\% of the total
neutrino emission, or $\sim 10^{51}$ erg.  As the estimate of the
total energy in the relativistic component of a GRB has come down in
recent years, the neutrino version of the collapsar model has become
more attractive. However, up to three orders of magnitude greater
energy is available from methods that more directly tap the
gravitational potential of the disk or the rotation of the black
hole. The neutrino version produces a hot, high entropy jet, while
some versions of the MHD models produce colder, Poynting flux jets
(\Sect{poynting}).

\begin{figure} 
\centerline{\psfig{file=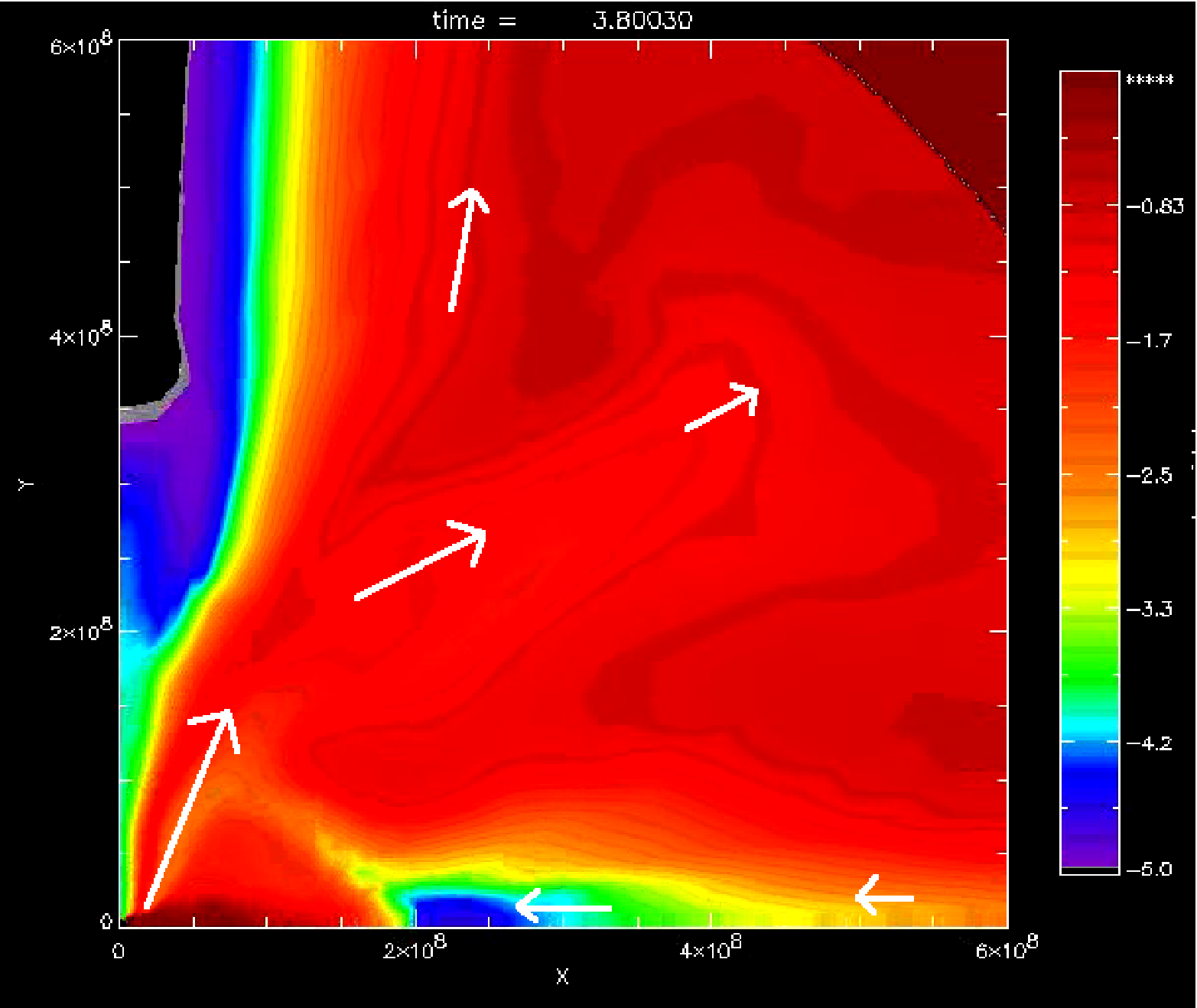,width=4in,angle=0}}
\caption{A ``wind'' of nucleons blows off the black hole accretion disk.
In the collapsar model, this wind is responsible both for blowing up
the star and producing the $^{56}$Ni to make it bright. In this
numerical simulation, the action of magnetohydrodynamical
instabilities in the disk is represented by a simple ``alpha-disk''
viscosity ($\alpha \approx 0.3$). The highest wind velocity (long
white arrow) is $\sim$20,000 km s$^{-1}$, and the mass in the wind is
a fraction ($\sim50$\%) of the 0.1 \Msun\ s$^{-1}$ accretion
rate. Matter flows in at the equator, is photodisintegrated, and
ejected as neutrons and protons.  Farther out these nucleons cool and
assemble to $^{56}$Ni. This is a separate phenomenon from the core jet
that makes the GRB. The figure is color coded by temperature with the
hottest (dark red) region being $\sim 4 \times 10^{10}$ K. The outer
radius of the figure is 6,000 km cm, so the inner disk of 300 km is
the dark part, in the lower left corner.
\citep{Mac99,Mac03}} \label{fig:niwind}
\end{figure}

In the collapsar model, the supernova and the GRB derive their
energies from different sources. The supernova, and the $^{56}$Ni to
make it bright, are produced by a disk ``wind'' \citep{Mac99,Mac03,Koh05}.
This wind is subrelativistic with a speed comparable to the escape
velocity of the inner disk, or about 0.1 c (Figure
\ref{fig:niwind}). If 1 \Msun \ accretes to make the GRB and half of 
this is lost to the wind, this is 10$^{52}$ erg.  The wind begins as
neutrons and protons in nearly equal proportions and thus ends up,
after cooling, as $^{56}$Ni. This nickel probably comes out in a large
cone (polar angle $\sim$1 radian) surrounding the GRB jet though it
might get mixed to other angles during the explosion.

In numerical simulations so far, the explicit MHD processes that drive
the wind have not been followed. An $\alpha$-viscosity is used instead
and the wind is driven by thermal dissipation. Thus the larger energy
(10$^{52}$ erg) of the supernova definitely has an MHD origin (if only
the disk instabilities responsible for its viscosity, \cite{Bal99})
while that of the GRB remains ambiguous.  There are also versions
of the collapsar model in which the black hole is not made promptly by
the failure of the initial shock to truncate accretion, but by fall
back in a supernova that explodes with insufficient energy for all its
matter to escape \citep{Mac01} and in pair instability supernovae at
high red shift \citep{Fry01}. Both of these variations probably give
transients that last longer than the typical value, 20 s, for ``long''
soft GRBs, but if they are to work, the energy source must be
MHD. Neutrino annihilation is too inefficient for the low accretion
rate in the fall back model and large black hole radius in the
pair-instability model.

It is a prediction of the collapsar, (though possibly the other
models as well), that the central engine remains active for a long time
after the principal burst is over, potentially contributing to the GRB
afterglow \citep{Bur05}. This is because the jet and disk wind are
inefficient at ejecting all the matter in the equatorial plane of the
pre-collapse star and some continues to fall back and accrete
\citep{Mac01}. 

Finally, the collapsar model attempts to explain the time structure of
GRBs and to produce the variable Lorentz factor necessary for the
internal shock model to function \citep{Pir05}. The jet, as it passes
through the star, is modulated by its interaction with the surrounding
matter \citep{Zha03}. That is, even a jet introduced with constant power in
the star's center emerges with a highly variable density and
energy at the surface. However, this interaction happens far from the 
central engine and would be present in any model where a relativistic
jet of radiation and matter must penetrate the star.

\subsubsection{Supranovae}
\lSect{supranova}

The maximum mass for a differentially rotating neutron star can be up
to $\sim$50\% larger than in the non-rotating case
\citep[T. Gold as cited in][]{Bla72,Mor04}. This gives baryonic maximum 
masses around 3 -- 3.5 solar masses, well above the largest iron core
masses expected in massive stars \citep{Woo02}. Uniform rotation causes
less of an increase, $\sim$20\%. The simplest version of the supranova
model \citep{Vie98} assumes that such a ``hypermassive'' neutron star
initially forms with a mass above the critical value for a slowly
rotating neutron star. The external star is blown away in an initial
supernova that makes the neutron star. Some time later, years in the
original model, dipole radiation slows the neutron star to the point
that it collapses to a black hole. For a soft equation of state with
an adiabatic index $\Gamma-4/3 \ll 1$, as much as 10 -- 20\% of the
mass of the collapsing neutron star avoids capture and goes into a
disk about the central hole \citep{Sha04,Due04}. This disk accretes
and a GRB jet is produced by MHD processes. The supernova evacuates
the near region of troublesome baryons that might contaminate the jet
and also provides a shell of heavy elements about 10$^{16}$ cm from
the burst.  The discovery of x-ray lines in the afterglows of some
GRBs provided support for this model \citep{Vie01}, but to the extent
that the lines themselves have become questionable (\Sect{lines}),
that motivation is less compelling. The disk, neutron-star combination
produced here is similar to that in models for short hard bursts
(\Sect{shorthard}) and to the extent that those are inherently less
energetic than long soft bursts, one wonders if there is sufficient
energy and duration for a long soft burst.

It is clear in the case of events like GRB 980425 and GRB 030329 that
the supernova and the GRB happened nearly simultaneously --- within a
few days of each other at most. Delays between several hours and
several months are also ruled out because neither the GRB jet nor its
emissions would escape the still compact supernova. Nevertheless the
model is not ruled out for those GRBs in which no supernova has been
observed, nor is it ruled out for situations in which the delay is
seconds rather than years. Since the critical mass for differential
rotation is considerably larger than for rigid rotation, there is a
range of masses for which simply enforcing rigid rotation (while
conserving angular momentum) will lead to collapse. \citet{Bau00}
estimate the magnetic braking time to be
\begin{equation}
\tau_B \sim \frac{R}{v_a} \sim 1 \left(\frac{B}{10^{14} \ {\rm G}}\right)^{-1}
\, \left(\frac{R}{15 \ {\rm km}}\right)^{-1/2} \, \left( \frac{M}{3 \ M_{\scriptscriptstyle 
\odot}}\right)^{1/2} \ {\rm s.}
\end{equation} 
It is unclear whether the outcome of an object experiencing such a
collapse is deformation, accompanied by gravitational radiation,
complete collapse, or collapse to a black hole plus a disk. In any
case, this might be a transition object along the way to the collapsar
model. Indeed, for the angular momentum that is invoked, a
hypermassive neutron star rotating at break up is a likely, though
often ignored initial stage in the collapsar model.

\subsubsection{Poynting flux or fireball?}
\lSect{poynting}

Related to the uncertainty in the birth of the GRB jet in the above
models, is a key uncertainty in the nature of the jet itself. Does it
consist of hot baryons, thermally loaded with radiation and pairs
greatly exceeding the rest mass \citep{Pir99,Mes02}, or is the ``jet''
characterized by large scale magnetic fields, dynamically dominant and
present from start (at the central engine) to finish (in the
afterglow)\citep{Mes97,Lyu01,Bla02,Lyu03b}? In the latter case the baryons
play little role and may as well be absent.

Numerical simulations \citep{Alo00,Zha04} show that relativistic
fireballs, given mild initial collimation, can pass through stars of
solar radius and exit the star with their large energy per baryon
intact. No such calculations exist yet for Poynting flux jets to show
that their ordered electromagnetic energy does not become thermalized
on the way out. This difficulty might be overcome once the jet has
bored a hole so that there is a low density (albeit optically thick)
line of sight to the center of the star. That is, a jet could initially
be a fireball and make a transition early in the burst to being
Poynting flux dominated.

The most important characteristic, observationally, of Poynting flux
models is that there are no internal or reverse shocks. In the model
developed by Lyutikov and Blandford, the GRB emission comes from
10$^{16}$ cm not 10$^{13}$ cm.  Poynting flux models have the capability
of producing large polarization \citep{Lyu03b} which might be a
diagnostic of the model. A strong early optical afterglow, as in GRB
990123, may also be easier to accommodate in Poynting flux models
\citep{Zha02,Fan04}. Poynting flux models also predict no high energy
neutrino flux accompanying the GRB, but do predict that GRBs could be
the site of ultra-high energy cosmic rays
\citep{Bla02}.

\subsection{PROGENITOR STARS}
\lSect{progenitor}

All GRB progenitors must lose their hydrogen envelope prior to death.
The radius of even a blue supergiant is several hundred light seconds
and the head of the jet which is to make the GRB travels significantly
slower than light while inside the star \citep{Zha03,Zha04}. One might
envision situations where a very asymmetric supernova might occur,
powered by the same central engine as a GRB, but a ~20 s gamma-ray
burst of the common variety is very unlikely. This is consistent with
the observation that the limited set of spectroscopic SNe associated
with GRBs are, so far, of Type I.  The progenitors must also be
massive enough and occur frequently enough to explain the observed
statistics. Finally, since only a small fraction of massive stars make
GRBs when they die, special circumstances must be involved.

\subsubsection{Mass}
\lSect{mass}

Single stars over about 10 \Msun\ on the main sequence are required in
order to make an iron core and collapse to a compact remnant. Stars of
still higher mass produce more massive iron cores and have greater
accretion rates onto that core once it collapses. Higher mass stars
also have greater gravitational binding energy outside the iron core
\citep{Woo02}. Both effects make an explosion more difficult.  
It has been speculated that above some critical mass, a black hole
forms before an outgoing shock is launched \citep{Fry99}, setting the
stage for the collapsar model. For a somewhat smaller mass, a black
hole could still form from fallback
\citep{Mac01}, though the accretion goes on at a much lower rate. 
For stars that do not make black holes, the faster rotation associated
with neutron stars resulting from the death of the most massive stars
\citep{Heg05} still makes them more promising GRB progenitors. The necessary
helium core mass to make a GRB is probably near 10 \Msun,
corresponding to a main sequence star of at least 25 - 30 \Msun, but
there are production channels that involve appreciably lighter single
stars \citep{Woo05}.

There are also binary channels for making long soft GRBs including a)
the merger, by common envelope, of two massive stars, both of which
are burning helium in their centers \citep{Fry05}; b) the merger of a
black hole with the helium core of a massive star \citep{Fry98,Zha01};
or c) the merger of a black hole and a white dwarf \citep{Fry99a}. If
the common envelope is completely dispersed and the mass of the merged
helium core is $\sim$10 \Msun \ or more, model a) gives a resulting
Wolf-Rayet star similar to the single star models. Cases b) and c)
result in black hole accretion with higher angular momentum and
produce longer bursts, probably longer than typical GRBs. The white
dwarf merger model would not give a supernova like SN 1998bw or 2003dh
and might not give the observed degree of concentration of GRBs in
star-forming regions.

\subsubsection{Rotation}
\lSect{rotation}

The role of rotation in ordinary SNe has long been debated
\citep{Hoy46,FH64,Leb70,Ost71}. GRBs aside, current models assign a
role ranging from dominant \citep{Aki03,Whe05,Ard05}, to important
\citep{Tho05}, and unimportant \citep{Fry02,Fry04,Sch04}.
No such ambiguity surrounds the role of rotation in making GRBs. It is
crucial in all current models (\Sect{grbmodels}).

Attempts to model the evolution of angular momentum in massive stars
to the point where their iron cores become unstable to collapse have
yielded uncertain results. There is general agreement that the
omission of magnetic torques leads to cores that do indeed rotate
rapidly enough to make a GRB \citep{Heg00,Hir04}. Indeed, one could
easily end up with the converse problem - too great a fraction of
massive stars would make GRBs. However, incorporation of approximate
magnetic torques \citep{Spr02} in single stars that evolve through a
red giant phase gives too little rotation for GRBs \citep{Heg05},
though just about the right amount for pulsars \citep{Ott05}. This
suggests either that the estimated torques are wrong or that special
circumstances are required to make a GRB.  

Recently, \citet{Woo05} and \citet{Yoo05} have discussed the possibility that
single massive stars on the high velocity tail of the rotational
velocity distribution function might experience ``homogeneous
evolution'' \citep{Mae87}, bypassing red giant formation
altogether. Such stars can die with very rapid rotation rates and
large core masses but only if the metallicity is low.

The possibility that GRBs are a consequence of binary evolution is
frequently discussed
\citep{Sma02,Pod04,Tut03,Tut04, Fry05,Pet05}, but the same caveats
apply. Unless the merger occurs well into helium burning, stars with
solar metallicity end up with iron cores that rotate too slowly to
make GRBs if the estimated magnetic torques are applied
\citep{Pet05,Woo05}.

In general, the rotation rate for WR stars is not well determined
observationally, but is expected, on theoretical grounds, to be rapid,
at least for low metallicity \citep{Mey05}

\subsubsection{Metallicity}
\lSect{metallicity}

Both the mass and rotation rate of potential GRB progenitors are
strongly influenced by mass loss.  Expansion of deeper layers to
replenish what has been lost from the stellar surface causes them to
rotate more slowly and ultimately this torque is communicated to the
core. Mass loss also makes the star lighter and easier to explode,
hence no black hole.  Thus if one wants a GRB, it is helpful if the
mass loss rate, especially during the WR phase of the evolution
leading up to the GRB, is small.  WR stars of low metallicity are
known to have smaller mass loss rates
\citep{Vin05}, scaling approximately as Z$^{0.86}$ down to
metallicities of 1\% solar (where Z refers here to the primordial iron
abundance, not the abundances of carbon and oxygen on the surfaces of
WC and WO stars). GRBs will therefore be favored in regions of low
metallicity \citep{Mac99} as has been observed in several cases
\citep{Fyn03,Pro04,Gor05,Sol05}. Reducing the metallicity to below
10\% solar will therefore possibly increase both the frequency and
violence of the outbursts.  In the collapsar model, more stars will
make black holes and those holes will accrete more matter. In the
magnetar model, more rapidly rotating neutron stars will be made. This
does not preclude the possibility of GRBs in solar metallicity stars
by some rare channel of binary evolution, or the estimates of magnetic
torques used in the stellar evolution models may be too big.

\subsubsection{Frequency}
\lSect{frequency}

\citet{Mad98} estimate that the core-collapse rate of all massive stars
from redshift 1 to 4 amounts to an observed event rate of 20 per 4 arc
minute square on the sky. Over the full sky this corresponds to about
5 SNe per second, a number that should approximately characterize the
observable universe. GRBs on the other are thought to occur throughout
the universe at a rate of about 3 per day (ref). Correcting for an
average factor of 300 in beaming, this means that, universe-wide, the
GRB rate is only about 0.2\% of the supernova rate.  Even allowing for
a large number of events in which a GRB-like central engine might
produce a supernova or an X-Ray flash without a bright GRB, this
suggests that GRBs are a rare branch of stellar evolution requiring
unusual conditions. If, however, the GRB rate is strongly metallicity
sensitive this fraction might increase with redshift. This is
consistent with observations that restrict the GRB rate to be no
greater than about 1\% the rate of core collapse supernovae
\citep{gal+05}.

\subsection{MODEL DIAGNOSTICS}
\lSect{diagnostics}

\subsubsection{The Supernova Light Curve and Properties}
\lSect{nucleo}

The fact that bright supernovae sometimes accompany GRBs offers a
powerful insight into the explosion mechanism. Type I supernovae of all
sorts are believed to be powered, at peak, by the decay of radioactive
$^{56}$Ni, and its daughter $^{56}$Co to $^{56}$Fe. $^{56}$Ni is only
made when matter with near neutron-proton equality (e.g., $^{28}$Si,
$^{16}$O, etc.) is heated to high temperature ($\gtaprx 4 \times 10^9$
K).  These high temperatures and the 6.077 day half-life of $^{56}$Ni
require that it be made in the explosion, not long before. Neither
merging neutron stars nor supranovae with a long delay are able to do
this.

In a spherically symmetric explosion, the production of $^{56}$Ni is
limited by the amount of ejected matter interior to a radius given by
$4/3 \pi r^3 a (4 \times 10^9)^4 ~ E_{\rm exp}$ where E$_{\rm exp}$ is
10$^{51}$ erg in an ordinary supernova, and perhaps 10$^{52}$ erg in a
GRB supernova. To make the 0.5 \Msun of $^{56}$Ni inferred from some
explosions, 10$^{52}$ erg must therefore be deposited in a time equal to
that required by the shock to go 10,000 km. Typical supernova shocks at
this radius move faster than 10,000 km s$^{-1}$, so the energy source
must radiate a power of $\sim 10^{52}$ erg s$^{-1}$ during the first
second of the explosion. This is far more power, at least during the
first second, than is required to make the GRB itself.

The GRB jet itself is relatively inefficient at making $^{56}$Ni
itself if it starts out with a small solid angle. This is both because
a small amount of matter is intercepted by the jet and because at very
high explosion energies the rapid expansion results in the production
of $\alpha$-particles, not just $^{56}$Ni. Still, one expects the
distribution of $^{56}$Ni in the ejecta to be very asymmetric and
concentrated along the rotational axis. Since a GRB observer is also
situated along this same axis, high velocities and large blue shifts
should be seen. The $^{56}$Ni may be rapidly mixed far out in the
explosion making it brighter, early on, than a more spherically
symmetric explosion. The deformation may also make the supernova light
mildly polarized.

An important issue is whether the amount of $^{56}$Ni and the
expansion rate are likely to be the same in all supernovae with GRBs
and XRFs, i.e., is the supernova a standard candle? Despite the
similarities between SN 2003dh and SN 1998bw (\S \ref{sec:spect}), the
answer probably is ``no''. One expects a large variation in the masses
and rotation rates of the progenitor stars, especially when
metallicity effects are folded in. Different stars will give different
rotation rates to their neutron cores, accrete different amounts of
mass into black holes of varying sizes, present different density
structures to the outgoing blast, etc. Supernovae expanding at a
slower rate will be fainter, even if they make the same amount of
$^{56}$Ni because their light will peak later after more decay and
adiabatic degradation. A ``supernova'' more than 10 times fainter than
SN 1998bw would be surprising, especially in the collapsar model,
which needs to accrete about a solar mass just to get the jet out of
the star. But it still might be explained either by pulsar models or
collapsars in which most of the energy came from black hole
rotation instead of disk accretion. On the other hand, a supernova
more than twice as bright as SN 1998bw would also be
surprising. Excepting pair instability supernovae, no such large
$^{56}$Ni mass has ever been verified in any supernova and it is
probable that a collapsar with such a vigorous wind would shut off its
own accretion. These limits are consistent with the observed spread in
supernova luminosities in XRFs \citep{Sod2005}.

\subsubsection{Unusual Supernovae and Transition Objects}
\lSect{orphan}

Supernovae are visible at all angles, last a long time, and do not
require relativistic mass ejection. There could therefore be a large
number of ``orphan'' supernovae in which the same central engine acts,
but which, for some reason, does not give an observable GRB. Perhaps the
jet had too much baryon loading or died prematurely. Perhaps a GRB
went off in another direction, or, for 20 seconds, we simply were not
looking at the right place.  Still the fraction of all
supernovae with the unusual characteristics of GRB-SNe is
small (\Sect{frequency}). It is also much easier to see the
GRB from a great distance, hence many GRB-SNe go undetected. A volume
limited sample of GRBs and broad line, Type I supernovae would
give interesting constraints on the beaming and jet break out, but so
far the data is too limited.

Specific cases include Type Ic supernovae SN 1997ef, 1997dq
\citep{Maz04} and SN 2002ap
\citep{Maz02,Yos03}. All show evidence for peculiarity and high energy.
Despite high velocity, a photosphere persists until late times.  SN
2003jd, also a Type Ic, shows evidence for two components of oxygen,
one at high velocity, one at low \citep{Maz05}. These observations are
best explained in a two component model \citep{Mae03,Woo03} in which
the the supernova has a high velocity along its rotational axes,
reflecting the activity of some jet-like energy source, and low
velocities in the equator.

Perhaps the clearest case so far of a transition supernova is SN
2005bf, \citep{Mae05,Tom05,Fol05}, a supernova that a) contained broad
lines of Fe II and Ca II, but only a trace of hydrogen, b) had two
distinct velocity components, and c) had two luminosity peaks, the
second being almost as bright as a Ia but occurring 40 days after the
explosion.  No spherically symmetric model with a monotonically
declining distribution of $^{56}$Ni is capable of explaining the
observations.  The explosion had to be an exceptionally massive helium
core to peak so late and yet make an unusually large amount of
$^{56}$Ni to be so bright then.  The explosion energy was also well
over 10$^{51}$ erg.  But the spectrum did not resemble so much that of
SN 1998bw as an ordinary Ib or Ic supernova.

All these supernovae so far are the explosion of WR stars of one sort
or another. It is expected on theoretical grounds that stripped down
helium cores will retain a higher angular momentum than those embedded
in red giant envelopes (\Sect{rotation}). Also, the asymmetry of the
explosion tends to be damped out in stars with extended envelopes
\citep{Wan01} and the high velocities in the helium core are tamped.
Still hyperactivity has been reported in Type II supernovae.  SN
1997cy \citep{Ger00}, the brightest supernova ever and a Type II, was
possibly associated with short hard burst GRB 970514. However with 4
month's uncertainty in the explosion date and difficulty making short
hard bursts in massive stars, this association is probably
coincidental. The high luminosity, attributed by Germany et al to 2.6
\Msun of $^{56}$Ni, may have been due to circumstellar interaction
\citep{Tur00}. Similar caveats apply to the identification of Type II 
SN 1999E with GRB 980910 \citep{Rig03}

\subsubsection{Energetics}

As was discussed in \Sect{characteristics}, the energy budget of a
GRB-SN can be broadly partitioned according to the Lorentz factor of
its ejecta. The supernova itself, as determined by its spectral line
widths, its nucleosynthesis, and some estimate of its mass, is a
measure of the non-relativistic ($1 < \Gamma < 1.005$) kinetic energy,
$E_{SN}$. The afterglow measures the energy in $\beta \Gamma \gtaprx$ 2
ejecta, $E_{Rel}$, and the GRB measures the energy of matter with $\Gamma
\gtaprx 200$, $E_{GRB}$. In general, one expects $E_{SN} > E_{Rel}$
and by definition, $E_{Rel} > E_{GRB}$, but there is no reason
that the ratio, $E_{SN}/E_{Rel}$, should be a constant from event to
event, and it probably isn't
\citep{mtc+04,skb+04}.

In the collapsar model, $E_{SN}/E_{Rel}$ measures the energy in the
disk wind compared to that in the jet. In the pulsar model, it
measures the ratio of prompt large-angle energy input to late time
input with small angles. In both cases, the answer could vary with
accretion rate, angular momentum, magnetic field strength, and stellar
mass. Conversely, observations of this ratio will ultimately constrain
these uncertainties in the models.

\subsubsection{Optical Spectroscopy of the Afterglow}
\lSect{spectrumag}

As light from the GRB and, later, the supernova passes through the
wind of the progenitor star, distinctive lines may be created that are
informative of the star's mass loss rate, wind speed, and composition
\citep{sgh+03,Mir03,Lan05}. Multiple components are seen with speeds
$\sim500$ km s$^{-1}$ and $\sim 3000$ km s$^{-1}$ The highest velocity
lines probably originate near the progenitor and reflect the WR wind
speed when it died, but lower speed lines are produced by a nebula
farther out that is either the fractured residual of a collision
between the WR wind and a previous red supergiant wind or radiatively
accelerated by the burst.  Observations so far are consistent with the
wind of a WR star, perhaps of class WC, but constrain the lifetime of
the WR progenitor to be shorter than expected \citep{Lan05}.

Some have suggested an origin as due to radiative acceleration of
dense clumpy nebular material by the GRB afterglow
\citep{mhk+02,sgh+03,rgs+05}. Then the absence of detectable
photoionization or deceleration places constraints on the
circumburst radii of the absorbing material \citep{mhk+02,sgh+03} and
possible companions to the progenitor \citep{swh+05,fdl+05}. Other
(less locally based) origins are possible, such as SNe remnants
\citep{lrcg02}, quasar absorption systems \citep{mhk+02}, or galactic
superwinds. Given the alternatives, and since only a handful of GRBs
have exhibited such blueshifted absorption, absorption spectroscopy
has yet to establish a definitive connection with the circumburst
environment. Still, with more frequent spectroscopic observations at
early times after a GRB, there is hope that the measurement of
time-dependent metal columns and velocities could provide important
diagnostics of the progenitors.

\subsubsection{X-Ray Lines}
\lSect{lines}

X-ray lines are potentially a powerful diagnostic of the supernova-GRB
combination. Lines from various elements have been reported in the
afterglows of at least seven GRBs using a variety of instruments
\citep[e.g.,][and references therein]{Pir00,Ree02,Pir05}. They are 
sometimes seen in emission about 10 hours after the burst with a
luminosity of 10$^{44}$ - 10$^{45}$ erg s$^{-1}$ for at least several
hours. Typically the lines are Fe-K$\alpha$, though the K$\alpha$
lines of Si, S, Ar, and Ca have also been reported. Unfortunately, the
statistical significance of these signals is not universally accepted
\citep{Rut03}. If real, the lines might be expected
in the supranova model, though other alternatives have been discussed
\citep{Mes01a,Kum03,Kal03} in which the supernova and GRB are simultaneous.

\subsubsection{GRB-SNe Remnants}

The supernovae that accompany GRBs may be hyper-energetic
compared with ordinary supernovae and are certainly accompanied by
jets. These peculiarities might manifest themselves in the remnant a
long time after the explosion is over. However, the jet energy is
probably less than the supernova energy, and initial asymmetries are
eroded once the explosion has swept up several times the initial mass
of the progenitor star. \citet{Aya01} estimate a time $\sim$5000 years
for this to occur, and, given the low event rate for GRBs, estimate a
remnant population $\sim$0.05 per galaxy. On the other hand,
\citet{Per00} and \citet{Rob03} discuss unusual SNR's that may have involved 
unusual energy or asymmetry. These observations may have other
explanations though - e.g., multiple supernovae. 

\subsubsection{Compact Remnants}

The black hole left in the collapsar model would be very rapidly
rotating (Kerr parameter $\sim$ 1), but the spin of an isolated black
hole is unobservable from far away. If the GRB occurred in a binary
system that somehow remained bound (admittedly a big {\sl if}) and
later became an accreting x-ray source \citep{Bro00}, measurement of
the mass and Kerr parameter \citep{Mid06,Sha06,Psa04} of the black hole
would both show large values. Indeed, measurements of Kerr parameters
in the range 0.5 to 1, even for black holes that may {\sl not} have
made GRBs, but do not seem to have been spun up by later accretion,
would strongly suggest that rotation is an important component of some
supernovae.

The object left in the millisecond magnetar model would be a very
magnetic neutron star (B$\sim 10^{15}$ G), with a still rapid rotation
rate and high luminosity, though probably not visible unless the burst
were relatively nearby. This would essentially be the birth of a
magnetar. Since the magnetic activity of such objects has been
associated with soft gamma-ray repeaters \citep[e.g.,][]{wt+05} and
since SGR activity might possibly be visible out to a distance of 70
Mpc \citep{Tan05}, nearby GRB sites, specifically that of GRB 980425 at
40 Mpc, might be monitored for repeated bursting activity.

\subsubsection{Nucleosynthesis}

\citet{Pru04a} and \citet{Sur05} find that the nucleosynthesis in the disk wind of
collapsars consists mostly of $^{56}$Ni for the relevant range of accretion
rates. The winds do not preserve the large neutron excess
characteristic of the inner disk because the outgoing nucleons capture
electron-positron pairs and neutrinos. If ``bubbles'' remove disk
material at an unusually low density and rapid rate, heavier nuclei
can be produced, even the $r$-process, but this is highly uncertain.

In the ideal case GRBs would greatly overproduce one or more nuclear
species not made elsewhere. \citet{Pru04b} have found that the wind
from collapsar disks can synthesize interesting large abundances of 
$^{42}$Ca, $^{45}$Sc, $^{46,49}$Ti, and $^{63}$Cu, but these same
species can be produced in ordinary supernovae \citep{Woo02}

\subsubsection{Afterglows and Density Gradients}
\lSect{afterglows}

The afterglow of a GRB in radio and x-ray is generally regarded as
coming from the external shock of the GRB producing jet as it
decelerates in the external medium. Breaks in the light curves of this
emission can yield information on the opening angle of the jet and
therefore on the actual total relativistic energy in the event
\citep{fks+01,Pir01,Pan02}.  In addition, the afterglow offers unique
insight into the mass loss history of the star just before it
exploded. If the metallicity is low, one expects mass loss rates much
smaller than for typical WR stars in our galaxy
(\Sect{metallicity}). 

It is important to note that radio emission from GRBs and Type Ibc
supernovae samples the mass loss during an epoch of stellar evolution
that is otherwise unobserved (and therefore not tightly
constrained). During the last several hundred years of their lives WR
stars over 8 \Msun are burning carbon and heavier elements in their
cores \citep{Woo02}. For a wind speed of 10$^8$ cm s$^{-1}$, this mass
loss determines the distribution of mass out to 10$^{18}$ cm wherein
all the afterglow is formed. The mass inside 10$^{15}$ cm, where the
burst itself gets made, reflects the last few months in the star's
life when it was burning oxygen and silicon. So long as the lass loss
rate depends only on the surface luminosity of the star, it will not
change much, for a WR star of given mass and metallicity, from helium
burning until explosion. The luminosity varies by only about 50\%.
But if these late stages are pulsationally unstable with a short
growth time, the mass loss could be quite different - perhaps higher.
The mass loss of WR-stars is also known to be clumpy \citep{Ham98},
and that could complicate the modeling.

In general, though, unless the mass loss is rapidly varying, which
seems doubtful in carbon burning, the density should scale as
$r^{-2}$. This scaling is consistent with radio observations of some
GRBs \citep{cl00,Li01,Pan02,pbr+02,gks+03}, but inconsiderate with others
\citep{cl99,kp03}. The latter is difficult to reconcile with the otherwise
successful paradigm that long soft GRBs originate from the deaths of
massive stars, but the complex interplay between the winds and the
interstellar medium could mask global wind signatures and even mimic a
constant density environment
\citep{wij01,clf04,rgs+05}.

\subsubsection{Gravitational Radiation and Neutrinos}

All models produce compact objects and require a lot of rotation and
thus predict a gravitational radiation signature of some sort
\citep{Fry01,Dav02,Kob03,Puta04,Putb04}. However, most of the models
are cylindrically symmetric. Perhaps the best opportunity would be
from the initial collapse that leads to the collapsar. The
proto-neutron star has more angular momentum than even a neutron star
rotating at break up and thus might pass through a highly deformed
stage before collapsing to a black hole \citep{Bau00}. But the
cylindrically symmetric exclusion of the excess angular momentum in a
disk is also a possibility that could greatly diminish the
gravitational radiation.

The neutrino burst from core collapse is not much brighter than in
ordinary supernovae and may even be fainter. Given the large distances
and soft spectrum, these neutrinos are probably not visible above 
the background. However, very energetic neutrinos can be produced by a
relativistic jet traversing a massive star \citep{Mes01,Raz04,Raz05}.

\section{THE FUTURE - A MYSTERY UNSOLVED}

\lSect{future}
 
While the observations of the last seven years have revealed an
exciting link between supernovae and the long-soft GRBs, it would be
premature to think that we understand either one of them very well. No
complete physical model currently exists for even the most common
variety of supernova. Indeed, one of the most important consequences
of the SN-GRB connection may be a better understanding of how massive
stars die.

Some specific diagnostics that might help with this were given in
\Sect{diagnostics}. Here we mention a few places where we think
significant progress could happen in the next decade. Some progress
will come simply from a larger sample of GRB-SN and from codes of
increased realism running on more powerful computers. Other advances
may require the development of space missions beyond Swift and
ground-based facilities that are only now in the planning stages. We
restrict our list to science specifically related to the SN-GRB
connection, not everything we want to know about, or can do with GRBs.

\begin{itemize}

\item{How variable --- in energy, mass, and luminosity --- is the
class of supernovae that accompany XRFs and GRBs?  We have taken the
position here that all of these high energy transients, except perhaps
the short bursts, are accompanied by stellar explosions of some
sort. Is that true? Were SN 1998bw and SN 2003dh unusually bright? Are
there any systematic differences in the supernovae that accompany
long-soft GRBs of different duration, energy, spectral hardness,
etc.?}

\item{Are GRBs favored by low metallicity? Do the average properties
of GRBs vary significantly (in their rest frame) with redshift? Since
mass loss decreases with metallicity, GRBs from high redshift might
preferentially come from more massive and more rapidly rotating
stars. This might reflect in the properties of the bursts and their
afterglows.}

\item{Pushing this to the extreme, can we use GRBs to study Population
III stars at very high redshift, including stars of much higher mass
than die as supernovae today? Bursts from redshift 10 -- 20 would be
both highly time dilated and severely reddened. A new mission or
mission strategy may be necessary that combines observations in the
infrared and hard x-ray.}

\item{What is the most common form of GRB in the universe? It is
possible that observations so far have been selectively biased to
more luminous events. Are events like GRB 980425 actually more
frequent than ``ordinary'' GRBs? More sensitive studies over a long 
period could eventually give, at least, a volume-limited local sample.}

\item{What is the is the relation between XRFs and GRBs? Are XRFs the
result of GRBs seen off axis, the result of jets that have lower
Lorentz factors at all angles, or something else? Observationally, it
will be important to see if the distribution of properties of XRFs and
GRBs is continuous from one extreme to the other. Theoretical models
are still primitive. Are both phenomena due to internal shocks or is a
mixture of internal and external shocks involved?} 

\item{Similarly, is there a continuum of events between core collapse
supernovae and GRBs, or are they two discrete classes of phenomena?
Rapid differential rotation in the core of a massive star when it dies
has been implicated as a necessary ingredient for GRBs.  Rotation may
play a role in producing all manner of unusual supernovae like those
mentioned in \Sect{orphan}, even those that have no GRBs. But is it
important in ordinary Type IIp supernovae?}

\item{How is the jet launched in a long soft GRB? Is the jet a
fireball or Poynting flux? This is largely an ongoing issue for theory
and simulation with important implications for active galactic nuclei
and pulsars as well as GRBs. There may be observational diagnostics,
however, in the polarization of afterglows and the strength of the
optical afterglow.}

\item{How long does the central engine operate? Is its power at late
times continuous or episodic? Recent studies with Swift have shown
some evidence in some bursts for substantial energy input continuing
long after the main burst is over. Variable late time energy input
could be a consequence of incomplete ejection of all mass in the
supernova which leads to continued accretion in the collapsar
model, though pulsar-based explanations are not ruled out.} 

\item{Do supernova-like displays ever occur with short hard bursts?
Present data suggest that they do not, but the exceptions should
continue to be sought.}

\item{In the longer time frame, neutrino bursts and gravitational
radiation may possibly yield the greatest insight into the nature and
activity of the central engine, since it is only in these emissions
that the central engine is directly observable.}

\end{itemize}

\section*{Acknowledgements}

This review has greatly benefited from discussions with many people
and presentations at conferences too many to mention by name. We are
particularly grateful, however, to Roger Blandford for a critical
reading of the manuscript and many useful comments; Alex Filippenko
for discussions of the characteristics of GRB-SNe; Alex Heger for
helping us understand the role of rotation in the advanced stages of
stellar evolution; Thomas Janka for critical comments on supernova
models; Chryssa Kouveliotou for discussions of GRB 980425 and GRBs in
general; Andrew MacFadyen and Weiqun Zhang for discussions of the
collapsar model and three of the figures used in the text; Tom
Matheson for his careful reading and comments, especially on GRB-SNe
properties; Bethany Cobb for comments; and Alicia Soderberg for many helpful comments on SNe and XRFs and a detailed critique of an earlier draft of this
manuscript. JSB also offers special thanks to Andrew Friedman, Dale
Frail, Shrinivas Kulkarni, and Robert Kirshner.


\end{document}